\documentclass[10pt, journal]{IEEEtran}
\IEEEoverridecommandlockouts
\usepackage{graphicx}
\usepackage{subfig}
\usepackage{amsmath,amsthm,amsfonts,amssymb}
\usepackage{cite}
\usepackage{bm}
\usepackage{bbm}
\usepackage{url}
\usepackage{array}
\usepackage{color}
\usepackage{multirow}
\usepackage{booktabs}
\usepackage[table,xcdraw]{xcolor}
\usepackage{enumerate}
\usepackage{enumitem}
\usepackage{epstopdf}
\usepackage{tabularx}
\usepackage{algorithm}
\usepackage[english]{babel}
\usepackage{algpseudocode}

\algrenewcommand{\algorithmiccomment}[1]{\hfill// #1}

\allowdisplaybreaks[4]
\begin{document}

\title{\makebox[\linewidth]{\parbox{\dimexpr\textwidth+0cm\relax}{\centering LLMind 2.0: Distributed IoT Automation with Natural Language M2M Communication and Lightweight LLM Agents}}}

\author{
Yuyang~Du$^*$,
Qun~Yang$^*$,
Liujianfu~Wang$^*$,
Jingqi~Lin$^\dagger$,
Hongwei~Cui,
Soung~Chang~Liew$^\ddagger$~\IEEEmembership{Fellow,~IEEE}
\thanks{Y. Du, Q. Yang, L. Wang, H. Cui and S. C. Liew are with the Department of Information Engineering, The Chinese University of Hong Kong, Hong Kong SAR, China (e-mail: \{dy020, yq020, wl024, ch021, soung\}@ie.cuhk.edu.hk). J. Lin is with The Hong Kong University of Science and Technology, Hong Kong SAR, China (email: jlindg@connect.ust.hk).}
\thanks{$^*$Y. Du, Q. Yang and L. Wang contribute equally to this work.}
\thanks{$^\dagger$This work was completed during J. Lin's internship at CUHK.}
\thanks{$^\ddagger$S. C. Liew is the corresponding author.}
\vspace{-2.5em}
}

\maketitle

\begin{abstract}
Recent advances in large language models (LLMs) have generated great interest in their applications for IoT automation and device management. However, centralized approaches struggle to scale across heterogeneous, large-scale systems. We present LLMind 2.0, a distributed framework that embeds lightweight LLM-empowered device agents and adopts natural language for machine-to-machine (M2M) communication. In LLMind 2.0, a central coordinator translates human instructions into natural-language subtask descriptions, which instruct distributed device agents to generate device-specific code locally based on their proprietary APIs. Using natural language as a unified medium overcomes device heterogeneity and enables seamless device collaboration. LLMind 2.0 integrates: 1) a timeout-based deadlock avoidance protocol that coordinates distributed subtask executions, 2) a retrieval-augmented generation (RAG) mechanism for precise subtask-to-API mapping, and 3) fine-tuned lightweight LLMs for reliable, device-specific code generation. Experiments in multi-robot warehouse operations and Wi Fi network deployments show LLMind 2.0’s improved scalability, reliability, and responsiveness compared to centralized baselines.
\end{abstract}

\begin{IEEEkeywords}
LLM agent, IoT automation, machine-to-machine communication, supervised fine-tuning
\end{IEEEkeywords}

\section{Introduction}\label{sec-I}
\IEEEPARstart{R}{ecent} breakthroughs in generative AI \cite{deng2025exploring} have sparked significant interest in applying large language models (LLMs) to enable efficient automation and device management \cite{4207411, 7128676, 10726709} in IoT systems with heterogeneous, collaborative devices \cite{cui2024llmind, xiao2024efficient, kok2024iot,qin2023toolllm, chen2024octopus, patil2023gorilla}. In these systems, a centralized LLM coordinator translates natural‑language instructions from human users into device‑executable code and then uses this code to control IoT devices via their APIs to fulfill the given task.

While effective for small-scale IoT systems, these approaches depend on a monolithic centralized coordinator and a code-based machine-to-machine (M2M) interface \cite{11264303,7128676}. This centralized design poses significant scalability challenges in larger, heterogeneous deployments, including:

\begin{enumerate}
    \item \textbf{API Management Complexity:} Proprietary APIs across diverse devices often come with device‑specific specifications and may even be implemented in different programming languages. As device diversity grows, this heterogeneity makes it progressively harder for the LLM to analyze tasks and coordinate with each device’s API, restricting system scalability.
    \item \textbf{Latency Bottleneck in the Coordinator:} The centralized coordinator is responsible for generating control scripts for all devices in the system. As the number of devices grows, the processing latency of the coordinator naturally becomes a bottleneck, reducing overall system responsiveness.
    \item \textbf{Reliability Challenges in Code Generation:} Ensuring the reliable generation of control scripts becomes increasingly difficult as both the diversity and the total number of managed devices scale up, raising concerns about the robustness and accuracy of the coordinator.
\end{enumerate}

Building upon LLMind 1.0 in \cite{cui2024llmind}, a representative centralized approach from prior work, this paper proposes \textbf{LLMind 2.0}, an IoT management framework that addresses scalability challenges by 1) adopting natural language as the M2M communication interface; 2) leveraging device agents empowered by lightweight LLMs to generate code. 

LLMind 2.0 differs significantly from recent approaches that use natural language as the M2M communication interface, offering greater adaptability across devices. Previous works \cite{cui2025towards,gao2025langcoop, bhatt2025uncap} map predefined driver commands to a fixed set of operations for vehicular control without generating new code (e.g., executing a pre-written script to move a car 10 meters by issuing the command “go ahead”). In contrast, LLMind 2.0 employs on-device code generation via a smart device agent, dynamically creating control scripts in real time. This enables the device's API to be utilized more flexibly, resulting in a more universal and adaptable framework.

\begin{figure*}[htbp]
  \centering
  \includegraphics[width=0.9\textwidth]{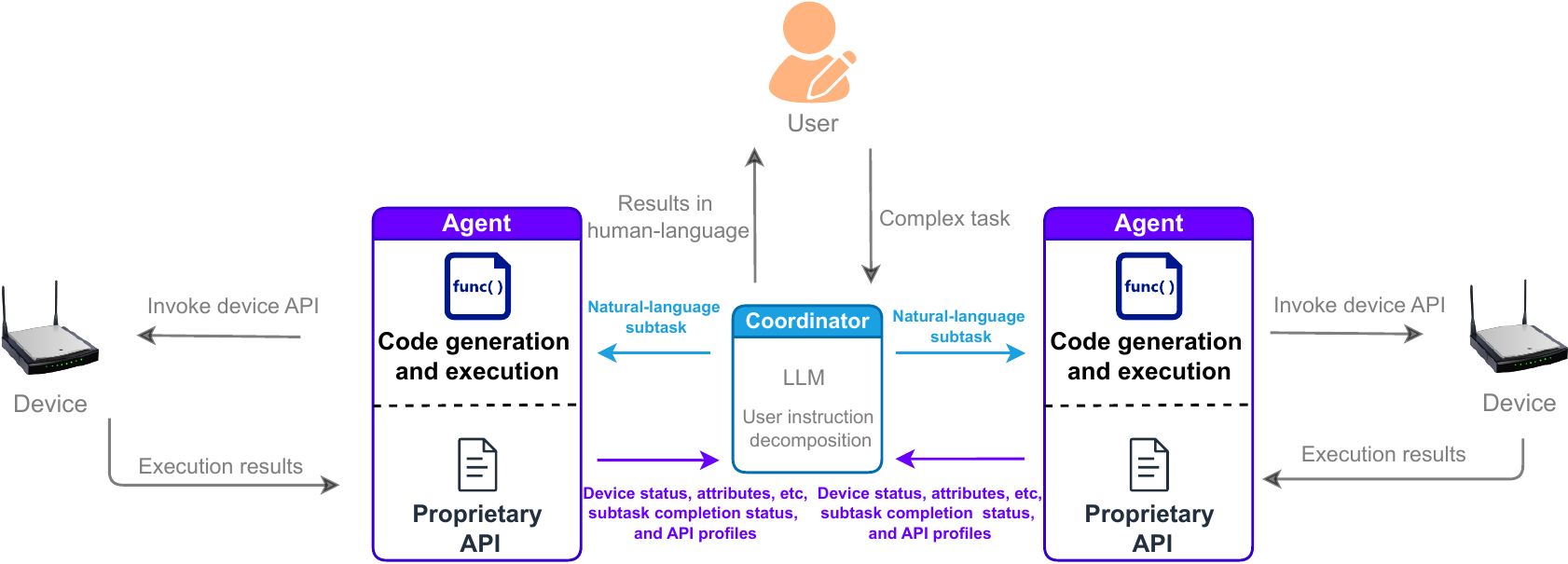}\\
  \captionsetup{font={small}}
  \caption{The LLMind 2.0 architecture addresses scalability challenges in managing diverse devices with proprietary interfaces. Device-specific agents, equipped with lightweight embedded LLMs, process the given natural-language subtask description and generate code employing their respective device-specific APIs.}
\label{fig:1}
\end{figure*}

As illustrated in Fig. \ref{fig:1}, this approach employs a conventional large-scale LLM as the central coordinator, tasked with translating human commands into a series of subtasks, each expressed in \textbf{natural language}. However, rather than directly generating executable code for each device, the coordinator transmits these natural-language subtask specifications to device agents. These agents, equipped with \textbf{lightweight LLM agents}, generate executable code employing their respective devices’ proprietary APIs.

By offloading code generation tasks to individual devices, the \textbf{distributed code generation} approach of LLMind 2.0 introduces several key advantages:

\begin{enumerate}
    \item \textbf{Enhanced Scalability and Adaptability:} With API‑based code generation shifted to device‑specific agents, the system reduces strain on the central coordinator as the number of devices grows, since it no longer needs to generate code. This novel design enables more effective system scaling. For subtask assignment, the coordinator only keeps a copy of the description of the devices’ API so as to understand their capability and how their behavior can be changed.  
    \item \textbf{Parallel Code Generation:} The distributed architecture allows multiple device agents to translate natural‑language subtasks into executable code simultaneously. This parallel processing significantly reduces system response latency    
    \item \textbf{Improved Reliability:} Each device‑specific agent handles its assigned subtask using only the proprietary APIs associated with that device. Compared with the conventional approach – where a centralized LLM generates code for every device – this agent‑based design naturally narrows the scope of generation and improves reliability. 
\end{enumerate}

In addition to the major advantages above, LLMind 2.0 also enables \textbf{Human‑Readable M2M Interactions}. Using natural language as the communication medium allows devices to interpret and share instructions in a human‑readable format. This improves the transparency of M2M interactions, enabling human operators to more easily monitor logs, trace interactions, and debug systems when needed.

Although the LLMind 2.0 framework offers many compelling benefits, its design and implementation introduce nontrivial challenges. Reliable operation in this distributed system requires both robust communication between the central coordinator and distributed device agents and accurate code generation. To address these challenges, we have developed the following techniques for LLMind 2.0:

\begin{enumerate}
    \item \textbf{Deadlock-Avoidance Protocol Design:} As a distributed system, LLMind 2.0 must handle asynchronous device operations and potential response delays and failures, which can disrupt workflows. In response to this, we design a fault-tolerant protocol that avoids deadlock with a timeout mechanism for coordinating device agents. See Section \ref{subsec-III(B)} for details.

    \item \textbf{Accurate Subtask-to-API Mapping:} Reliable code generation at the agent begins with selecting the correct API. To ensure each device invokes the appropriate function for a given subtask, we implement a retrieval-augmented generation (RAG) \cite{wang2025cellular, gao2023retrieval, qi2025verirag} framework. Before generating code, the device agent retrieves the most relevant API from its device-specific knowledge base with a matching module. Section \ref{subsec-III(C)} details the design of this module.
    
    \item \textbf{Lightweight LLM Fine-Tuning for API Input-Parameter Extraction:} The next challenge for the agent’s reliable code generation is extracting accurate API input parameters from the natural-language description. To provide the device agent with error-free API input values, we fine-tune a lightweight LLM for accurate attraction of API input-argument values from the given subtask description. Section \ref{subsec-III(D)} presents the design of this LLM-empowered module in detail.
\end{enumerate}

For system validation, we implement LLMind 2.0 and conduct the following experiments:

\textbf{Experiment 1:} We simulate an unmanned warehouse where a manager remotely directs mobile robots to collaboratively locate vacant shelf slots. We compare the distributed scheme of LLMind 2.0 with the centralized approach in LLMind 1.0 \cite{cui2024llmind}, evaluating the execution of multi-robot tasks across heterogeneous robot platforms.

\textbf{Experiment 2:} We deploy a real‑world system with multiple IoT devices connected via WiFi. The system tackles two networking tasks: 1) QoS tuning via Enhanced Distributed Channel Access (EDCA) \cite{chang2005ieee} and 2) interference detection and mitigation. The results highlight LLMind 2.0’s distributed intelligence, with devices completing both tasks through cooperative and competitive behaviors.

The remainder of this paper is organized as follows. Section \ref{sec-II} discusses the limitations of conventional centralized approaches, using LLMind 1.0’s performance in Experiment 1 as a representative case study. Section \ref{sec-III} details the design principles and implementation of LLMind 2.0. Section \ref{sec-IV} validates LLMind 2.0 by 1) comparing its performance in Experiment 1 against LLMind 1.0’s results in Section \ref{sec-II}, and 2) analyzing cooperative and competitive behaviors in Experiment 2. Finally, Section \ref{sec-V} concludes this work.

To facilitate reproducibility and support further research, we release agent building details at: \textcolor{blue}{\url{https://github.com/1155157110/LLMind2.0}}.

\section{Motivations: Warehouse Case Study}\label{sec-II}
To better understand the limitation of the conventional approaches with code-based M2M communication interface, consider a case study of an unmanned warehouse with fixed shelves and mobile robots. For this case study, we adopt  LLMind 1.0 \cite{cui2024llmind} as a representative example of a conventional approach.

As emphasized in the introduction, device heterogeneity is a common challenge in IoT scenarios. In this case study, the robots are from different manufacturers with distinct API interfaces. A warehouse manager working remotely wants to know if there is a vacancy for new goods and issues a simple instruction: “\textit{Please check if there are vacant spaces on the shelves.}”

Conventional unautomated IoT systems \cite{10693298,10339255,10810266} require human planning and programming – e.g.,  assigning tasks to robots and manually programming each robot to execute its assigned task for the scenario in this case study. IoT automation systems empowered by LLMs, such as LLMind 1.0 \cite{cui2024llmind}, reduce the need for manual intervention through LLM-based task planning and code generation. 

However, centralized code-generation approaches like LLMind 1.0 face scalability challenges. Specifically, such systems generate code as the M2M interface between the central coordinator and a mobile robot. Due to device heterogeneity, the control script for one robot may not work for another, requiring the centralized coordinator to generate separate, device-specific control scripts for the robots. 

\begin{figure}[htbp]
  \centering
  \includegraphics[width=0.5\textwidth]{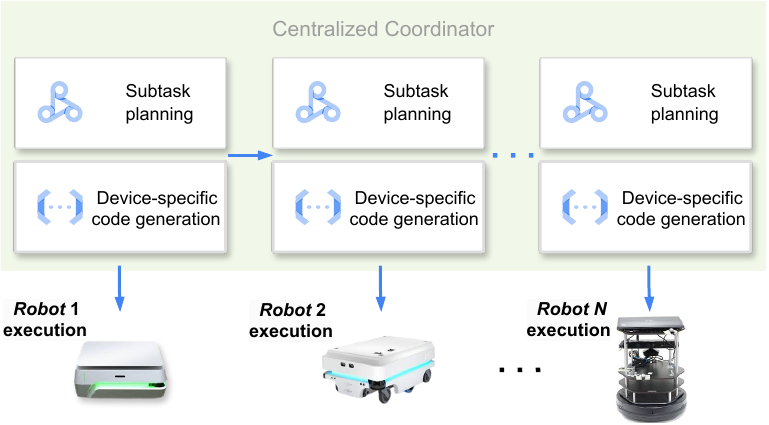}\\
  \captionsetup{font={small}}
  \caption{The processing pipeline of the centralized coordinator in LLMind 1.0 when given the shelf vacancy search task.}
\label{fig:2}
\end{figure}

As illustrated in Fig. \ref{fig:2}, the centralized coordinator may use a serial processing pipeline to orchestrate mobile robots in the search for shelf vacancies: 1) plan a task for Robot 1; 2) generate code for Robot 1; 3) repeat steps 1) and 2) for each subsequent robot.

We now describe how LLMind 1.0 processes this task to illustrate the limitations of centralized approaches, using GPT4 for the cloud realization of the coordinator. After receiving the task, the system coordinator analyzes the given task and determines how to allocate the available robots to inspect the shelves. Subtasks are distributed among the robots based on the number of robots available in the system. Let the number of fixed shelves and mobile robots be $M$ and $N$, respectively. If $N<M$, each robot will receive a subtask, with some robots required to check multiple shelves; if $N\geq M$, only $M$ out of $N$ robots will be activated, and each activated robot will be tasked with checking one shelf only.  The specifics of each subtask depend heavily on the capabilities of individual robots (such as their hardware and API interfaces). However, the general process follows these steps: 1) the robot moves to a designated location that allows it to clearly view the shelf, 2) the robot captures an image of the shelf, 3) the robot uses AI tools -- either deployed locally or cloud based -- to analyze the shelf for vacancies, 4) the robot reports the state of a shelf to the coordinator, and 5) the robot moves to another shelf and repeats the above operation if it is tasked with inspecting multiple shelves.

\begin{figure}[htbp]
  \centering
  \includegraphics[width=0.45\textwidth]{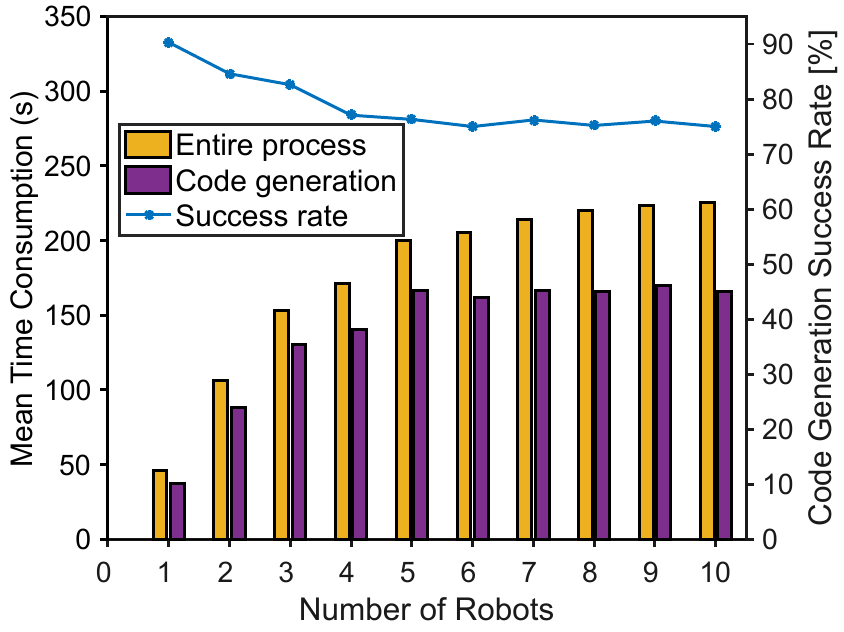}\\
  \captionsetup{font={small}}
  \caption{Reliability and latency performance of LLMind 1.0 for the shelf vacancy detection task. The figure presents the success rate of generating accurate control scripts for all involved robots, along with the code generation latency and the “entire process” latency, including the time for task planning and code generation. Note that we focus on the bottleneck at the coordinator to highlight the scalability limitations of the centralized code generation approach. The above latencies do not include the times required for the robots to physically move to the shelves, inspect the shelves, and report the statuses to the coordinator.}
\label{fig:3}
\end{figure}

Fig. \ref{fig:3} shows the performance of LLMind 1.0 in the warehouse task, where $M=5$ and $N$ varies from $1$ to $10$. As N increases from $1$ to $5$, the centralized code-generation approach faces increased task complexity as the number of heterogeneous devices grows, resulting in a lower success rate and longer latency in the code generation process. When $N>5$, although there is no need to generate control scripts for inactive robots, the latency and reliability of the system remain problematic. Fig. \ref{fig:3} also shows that code generation accounts for a large portion of the total processing time in the LLM coordinator.

Through the above case study, we highlight once again the scalability problems of LLMind 1.0-like system with centralized script generation and code-based M2M communication interface: the increased task complexity when diverse devices are involved, as well as resulting issues in latency and reliability. Section \ref{sec-III} addresses these problems, and Section \ref{sec-IV} validates our designs.

\section{System Design}\label{sec-III}
\subsection{Design Principles and System Overview}\label{subsec-III(A)}
The observations in Section \ref{sec-II} lead to the following key design principles underlying LLMind 2.0: 

\textbf{Communication-Coordination Protocol between Coordinator and Devices:} LLMind 2.0 introduces a distributed communication protocol that decentralizes intelligence across individual devices. By shifting code generation to device agents, the system enables concurrent code generation across distributed devices. This significantly reduces processing latency, which is particularly advantageous in Internet of Things (IoT) environments where real-time operation is critical. A primary challenge in designing this protocol is avoiding operational deadlocks. Technical solutions to address this issue are detailed in \textbf{Subsection B}.

\textbf{Natural-Language M2M Communication Interface:} In the distributed framework proposed, traditional code-based communication interfaces between the coordinator and devices are replaced with human language descriptions. This shift simplifies M2M communication, reducing the workload on the coordinator and enabling it to focus on critical tasks such as task decomposition and subtask orchestration. 

\begin{figure}[htbp]
  \centering
  \includegraphics[width=0.475\textwidth]{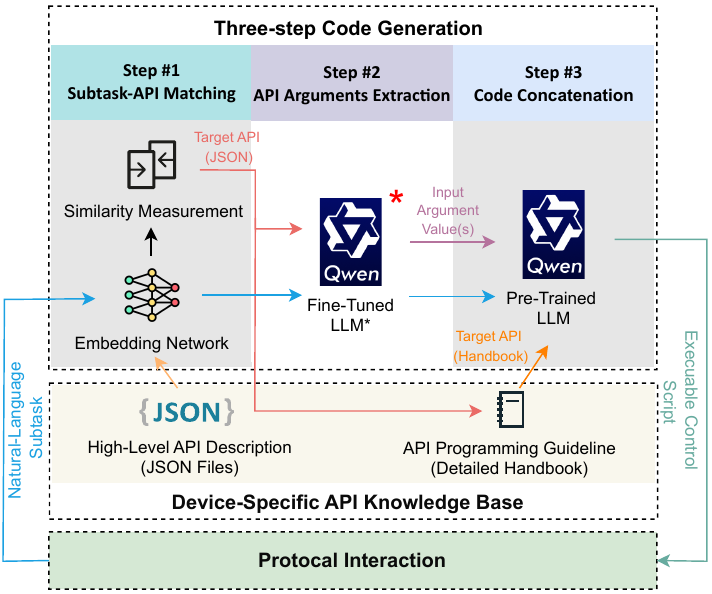}\\
  \captionsetup{font={small}}
  \caption{The framework of the proposed device agent.}
\label{fig:4}
\end{figure}

To realize this natural-language-based M2M Framework, LLMind 2.0 must achieve accurate code generation at distributed devices, even if the computational resource of an IoT device can only support lightweight LLMs available with much fewer parameters compared to cloud-based models available in centralized approaches.  Our solution for this critical challenge, as illustrated in Fig. \ref{fig:4}, is a three-step pipeline that decomposes the natural-language-to-code transformation task into easier steps:

\textbf{Step 1 (Subtask-to-API Matching):} A matching module within the agent system matches the given subtask, specified in natural language, with device-specific APIs to identify suitable API functions needed for completing the given subtask.

\textbf{Step 2 (Language Analysis for Input Extraction):} An argument-value extraction module within the agent system, which is empowered by a fine-tuned lightweight LLM, analyzes the subtask’s natural language description to extract the required input values for the API function’s arguments.

\textbf{Step 3 (API Call Completion):} Another lightweight LLM within the agent generates code based on the information obtained from Steps 1 and 2, allowing the agent to complete the API call on the associated device. 

\textbf{Subsection C} explains how the matching module in Step 1 maintains accurate subtask-to-API mapping under the agent’s RAG framework. The functionality of the matching module is highlighted as follows: a device agent has access to a local database containing information on device-specific APIs and detailed programming guidance for each API. Different devices may have different APIs, and LLMind 2.0 handles this heterogeneity by using the matching module to retrieve the most task-relevant API from the device-specific database for later RAG-based code generation.

\textbf{Subsection D} describes how a lightweight LLM is fine-tuned to fulfill the task in Step 2, i.e., reliably extract input values for API arguments from subtasks described in natural language. This is a common capability required of all device agents. Thus, all device agents utilize the same fine-tuned lightweight LLMs, supplemented by different RAG databases to address the heterogeneity among them.

Step 3 is positioned as the final code generation process within the RAG framework. In this step, the agent system leverages another lightweight LLM (different from the fine-tuned LLM in Subsection D) to generate executable control scripts based on 1) the relevant snippet of the programming guide associated with the target API found in Step 1; 2 ) API input-argument values extracted in Step 2, and 3) the user’s original language input. We note that Step 3 is a classic RAG-based code generation process in essence. Since this mature technique has been well studied in existing literature \cite{bassamzadeh2024comparative,nair2025prompts}, this paper does not dive into details therein \footnote{Instead of the RAG-based code generation and its implementation, we highlight that contributions of this paper lie in the design of the three-step code generation framework for LLMind 2.0 agent (illustrated in Fig. \ref{fig:4}), two vital modules tailored to the framework (detailed in Subsections C and D),and the deadlock-avoidance protocol to smoothly run the generated code (detailed in Subsections B).}.

\textbf{Device Handbook Information Provided to the Coordinator and Device Agents:}
A point that requires clarification for a more comprehensive system overview is the location and type of API information provided within the three-step code generation pipeline. In systems where a central controller is tasked with generating code for all devices, such as LLMind 1.0, each device’s API handbook is included in the prompt of the coordinator during its code generation. This allows the coordinator LLM to follow the detailed programming guidance of each device when generating the associated code. 

However, with the shift in LLMind 2.0 in delegating code generation to device agents, the coordinator now only requires brief descriptions of a device’s API.  These descriptions provide the coordinator with an understanding of the device’s capabilities to compose and allocate subtasks. There is no longer a need to store full API programming guidance at the coordinator, as lengthy device-specific materials pose scalability challenges (e.g., overwhelming the coordinator LLM with information) as the number of devices increases.

In LLMind 2.0, apart from detailed programming guides (see Fig. \ref{fig:A2} in Appendix A for example), API information can also be structured into a unified JSON format for efficient storage. The JSON-based API description, as illustrated in Fig. \ref{fig:5} and Fig. \ref{fig:A1} in Appendix A, includes essential fields such as the API name, inputs, outputs, and a brief functional description. The LLMind 2.0 coordinator is given JSON-formatted API information to facilitate composing natural-language subtasks. Such API information of all devices under the coordinator’s control is given as part of its system prompt.

\begin{figure}[htbp]
  \centering
  \includegraphics[width=0.49\textwidth]{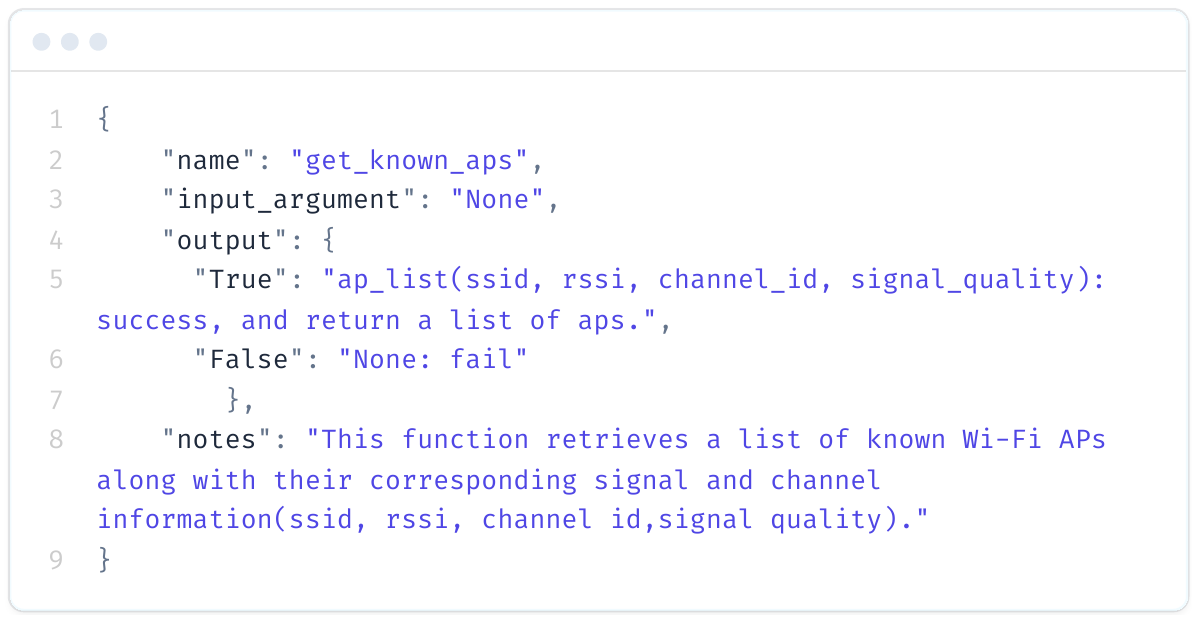}\\
  \captionsetup{font={small}}
  \caption{An example of a device API function in JSON format. The API function get\_known\_aps searches for nearby WiFi access points (APs) and returns a list of known APs. This function requires no input argument.}
\label{fig:5}
\end{figure}

On the agent side, as illustrated in Fig. \ref{fig:4}, the JSON-formatted API information for a device is provided to the agent during the subtask-API function matching process to facilitate easier and more efficient matching. Based on the matching result, the agent retrieves the relevant snippet of the programming guide for the target API to use in Step 3. Unlike conventional code generation in centralized systems, only a specific portion of the API handbook is retrieved to simplify the agent’s analysis, reducing the complexity of the task. 

\subsection{System Protocol Designs}\label{subsec-III(B)}
This subsection describes the interaction protocol between the coordinator and device agents. Specifically, we design a fault-tolerant periodic protocol with time-outs to avoid potential deadlocks. Before delving into the details, we give an overview of the coordinator-agent interaction:

1.	The coordinator periodically polls and collects device data from each device agent, including: 1) device’s current status; 2) device’s attributes, such as the latest hardware or register configurations; and 3) the completion status of the subtask running on the device. 

\begin{figure}[htbp]
  \centering
  \includegraphics[width=0.475\textwidth]{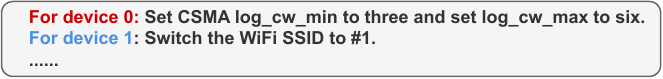}\\
  \captionsetup{font={small}}
  \caption{Examples of subtasks specified in natural language.}
\label{fig:6}
\end{figure}

2.	Upon an incoming human command, the coordinator decomposes the associated complex task into multiple subtasks specified in natural language for device executions according to the latest device data (see Fig. \ref{fig:6} for an example of the natural language specification for devices).\footnote{LLMind 2.0 follows the design of LLMind 1.0 when decomposing a complex user task into multiple subtasks, but it turns to natural language for the description of each subtask and shifts the language-code transformation to device agents.}

3.	The coordinator distributes subtask specifications to device agents, who then generate device-specific code for execution by the respective devices.

We next move to the details of the interaction protocol. 

\begin{figure}[htbp]
  \centering
  \includegraphics[width=0.475\textwidth]{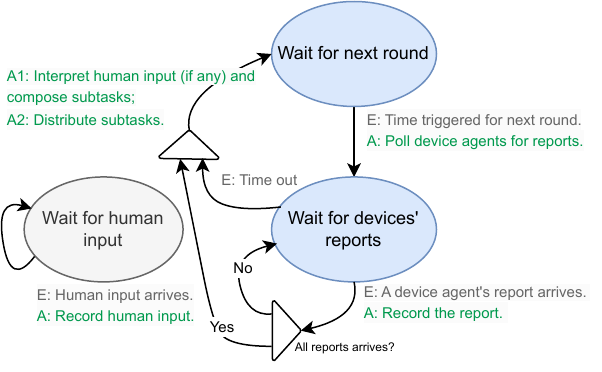}\\
  \captionsetup{font={small}}
  \caption{State transition diagram of the coordinator.}
\label{fig:7}
\end{figure}

Coordinator Protocol: Fig. \ref{fig:7} shows the coordinator’s state transition diagram with two threads of execution. The ellipses are the states, E represents the event that triggers the exit from a state, and A is the action performed by the thread between the transition from one state to another state. When a human user issues an instruction, the first thread records the input for later action(s) to be performed by the second thread. As long as there is an instruction from a human, the thread will proceed. This thread will never get into a deadlock situation where it is waiting indefinitely for a response from other threads because other threads do not report to this thread. 

In the second thread, the coordinator periodically polls device agents for reports, ensuring it regularly exits the “wait for next round” state and will not deadlock in this state indefinitely. The reports from a device agent include device status, the device’s most updated configurations, and subtask completion status. The device status indicates whether the device is operating normally, while the subtask completion status indicates whether the assigned subtasks have been completed, ongoing, or not executable. 

After the poll, the thread transitions to the state “wait for devices’ reports,” where it awaits the agents’ reports to arrive. As shown in Fig. \ref{fig:7}, once reports from all polled device agents have been received, the coordinator interprets them, checks for any new instructions from humans, and, if necessary, composes and distributes new subtasks for execution by the device agents. 

Note that an agent’s report may arrive late or fail to arrive entirely due to malfunctions. Nevertheless, as illustrated in Fig. \ref{fig:7}, a time-out mechanism ensures that the thread exits the “wait for devices’ reports” state and transitions to the “wait for next round” state once the time-out is reached. This mechanism guarantees that the thread will not deadlock in the “wait for devices’ reports” state, even if some agents or their devices malfunction and fail to return reports.

Regardless of whether all or only some reports are received, the coordinator will eventually proceed to interpret and act on the returned reports. Afterward, if there are pending human instructions, it composes the corresponding subtasks and distributes them to the agents. The coordinator concludes a round of operation once all subtasks have been assigned.

\begin{figure}[htbp]
  \centering
  \includegraphics[width=0.49\textwidth]{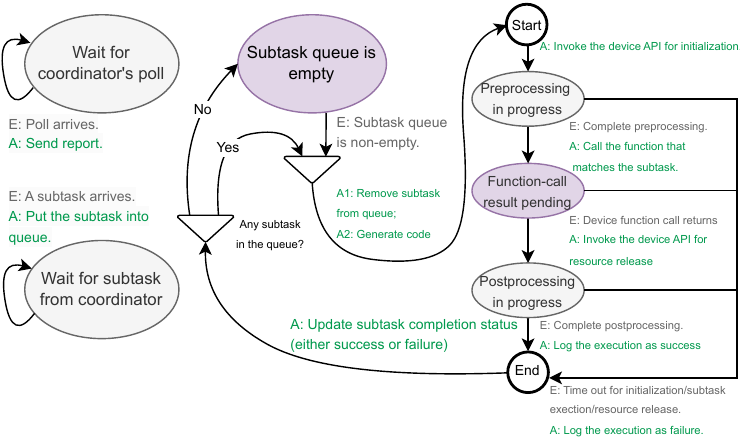}\\
  \captionsetup{font={small}}
  \caption{State transition diagram of the device agent.}
\label{fig:8}
\end{figure}

\textbf{Device Agent Protocol:} Fig. \ref{fig:8} shows the state transition diagram of a device agent consisting of three threads of execution. The first thread sends the report to the coordinator upon receiving the coordinator’s poll. The second thread pushes a natural-language subtask from the coordinator to a queue for access by the third thread. The two threads do not get deadlocked because the coordinator periodically polls device agents and assigns subtasks to them when a request comes from a human.

If the subtask queue is not empty, the third thread retrieves a subtask from the queue, interprets it in the context of the device API, and generates the corresponding executable code (as detailed in Subsections C and D). It then uses the generated code to invoke the appropriate initialization API function to prepare the device, followed by calling the API function associated with the subtask. Finally, it transitions to the state "Function-call result pending" and waits for the device to report the result. When the function call returns, the third thread proceeds to post-processing, invoking the device API to release resources and log essential information about subtask execution. The result of subtask execution is logged as success after the completion of postprocessing.

Note that the above three states (i.e., preprocessing, function-call result pending, and postprocessing) include a timeout mechanism to prevent deadlocks in case the device malfunctions or the invoked API fails to return a result. If any timeout happens in the three states, we log the result of subtask execution as a failure.

As shown in Fig. \ref{fig:8}, once the device completes post-processing, the third thread updates the subtask’s completion status and checks whether there are any outstanding subtasks in the queue. If there are, the thread repeats the above actions. Otherwise, it waits until the queue is filled with a new subtask before resuming its operations.

We note that the subtask queue operates with a single-cache mechanism: only the latest subtask is retained in the queue. If a new subtask arrives while another subtask is still in the queue, the new subtask supersedes the existing one, which is then dropped. The underlying idea is that the absence of execution results for the queued subtask, as observed by the coordinator, implies that the subtask has not been executed. In such cases, the coordinator may generate a new subtask to replace the one in the queue, accounting for the possibility that the device agent might have failed to execute the subtask on time. This failure could result from the device agent's slow processing speed, delayed response from the device itself, or other factors. 

This mechanism is designed to prevent deadlocks and redundant execution of subtasks. In essence, the current protocol operates under a best-effort assumption, where the execution of subtasks by device agents is not guaranteed. Furthermore, even when subtasks are executed, their results may not always be satisfactory. While this paper focuses on this specific interaction protocol between the coordinator, device agents, and devices, alternative protocols beyond the one explored here are certainly possible.

\subsection{RAG-based Mapping of Subtasks to Device API} \label{subsec-III(C)}
The complete detailed description of the device's available APIs, such as the user manual or developer handbook, can, in principle, be integrated into the agent system as an external knowledge base. However, a comprehensive developer handbook may include descriptions and details irrelevant to the specific subtask assigned to the device, potentially overwhelming the device agent, especially when implemented on a lightweight LLM.

To address the complexity issues, we implement a matching process to accurately identify the appropriate API function for a given subtask, enabling the retrieval of the relevant portion of the handbook. Specifically, we developed an RAG  matching module within the device agent to handle the mapping problem. The RAG module measures the similarity between subtask descriptions and API functions in the handbook to identify the most relevant function required for the subtask.

\begin{figure}[htbp]
  \centering
  \includegraphics[width=0.4\textwidth]{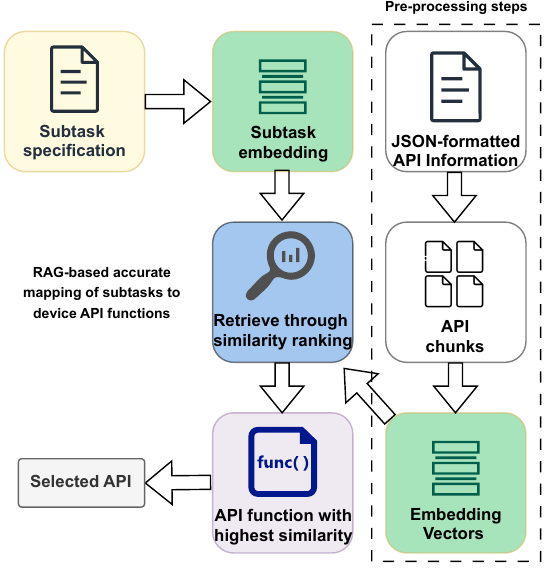}\\
  \captionsetup{font={small}}
  \caption{The RAG-based approach involves encoding subtask specifications and device API functions into sentence embeddings for semantic similarity matching.}
\label{fig:9}
\end{figure}

Fig. \ref{fig:9} depicts the RAG approach. As pre-processing steps, we build an embedding vector for each device API as follows. We start by extracting the functional description of an API from its JSON-formatted API information file (see Fig. \ref{fig:5} for the JSON-formatted API information) to form multiple RAG chunks, with each chunk representing an API of the device. For example, the API functional description for “adjust\_camera\_angle", which is listed as an example in Fig. \ref{fig:A1} of Appendix A, is “this function adjusts the camera's angle to a specified value, helping the robot to better scan areas or shelves”. We then encode the functional description into a vector embedding using a pre-trained Transformer model (we use SentenceTransformer \cite{SentenceTransformer} in our system building).

Recall that the same JSON-formatted API information of devices is also provided to the coordinator (see discussion at the end of Section \ref{subsec-III(A)} to help it to compose subtasks in natural language. Thus, we would expect the coordinator’s subtask description to lean heavily on the API information, especially the API function descriptions (e.g., a subtask description matching the “adjust\_camera\_angle” function in the example above might be “change the camera angle to 30 degrees”). 

Thus, during operation, when a subtask arrives, the device agent encodes it into a sentence-vector embedding using the same transformer network, SentenceTransformer \cite{SentenceTransformer}. The agent then computes the cosine similarities between the subtask embedding and each of API embeddings, retrieving the API function with the highest similarity as the best-match API of the current subtask. 

\subsection{API Function Input-Argument Values Extraction with the Fine-tuned LLM} \label{subsec-III(D)}

With the well-matched API function obtained in Step 1, we now move to Step 2 – extracting the values for the input arguments of the API function from the subtask’s natural-language description. 
distillation

As highlighted in Fig. \ref{fig:4}, inputs of Step 2 include JSON-formatted brief API information and the subtask description, while the outputs are specific values for API input arguments needed for completing the API function call. For example, let us consider \textit{“AC\_Setup (INT Temperature, INT FanSpeed)”}, a temperature and Fan Speed control API of a smart air conditioner, and a given subtask description is \textit{“Set the AC temperature to 26 and adjust the fan speed to level 2”}, the desired argument values should be $(26, 2)$.

Although Step 2 looks simple, our experiments on lightweight LLMs with no larger than 7B parameters indicate that these models frequently fail to precisely identify the API input-argument values from the subtask description (see Fig. \ref{fig:11} in Section \ref{sec-IV} for details). This is likely due to the lightweight model’s limited ability compared with powerful cloud-based LLMs such as GPT4. 

We address this problem by fine-tuning a Qwen3-0.6B model (a lightweight LLM selected to represent typical low-cost IoT devices \cite{yang2025qwen3}) using GPT4-generated data. The approach, which is in essence a knowledge distillation of GPT4, allows us to enhance the lightweight model's performance on a specific task to a level comparable to that of the powerful cloud-based model \cite{wang2025rephrase, zhang2025sa}.

Following the input-output requirement outlined above, we aim for three elements for each GPT-4 generated labelled data instance: 1) JSON-formatted description of an API function, 2) subtask description, and 3) desired API function’s input-argument values. To this end, we conducted GPT4-based LLM role playing as follows: 

1. We first ask GPT4 to play the role of a device manufacturer, creating API functions of imagined devices and generating a JSON-formatted API brief for each function. To reflect real-world scenarios, the generated API functions include varying numbers of arguments. Specifically, GPT-4 is instructed to randomly select between one to four arguments for each API function during the data generation process. For each imagined device, we generate 40 APIs.

2. For each generated API function, we then ask GPT4 to play the role of central coordinator and generate a subtask description that intends to use the given function.

3. Finally, we ask GPT4 to play the role of the argument-value extraction module in Step 2 and produce the expected argument-value extracting outcome when given the JSON-formatted API brief and the subtask description. 

Details of the fine-tuning process are as follows: 

1. The data construction process considers 1,200 devices, which results in 48,000 data instances in the API dataset. We randomly choose 90\% of imagined devices and use their associated APIs for model training (i.e., 1,080 devices with 43,200 data instances), while the remaining 10\% data are used as the testing data (i.e., 120 devices with 4,800 data instances).

2. With the GPT4-generated data, we follow the input and output requirements of Step 2 (detailed at the beginning of this subsection) to fine-tune the Qwen3-0.6B model.

3. LoRA fine-tuning \cite{sun2022recent} was applied to the Qwen3-0.6B model. For detailed configurations of the LoRA adaptation, we set LoRA\_rank=32, LoRA\_alpha=32, and LoRA\_dropout=0.1. Meanwhile, we have batch\_size=2, learn\_rate=2e-2, and we apply the 8-bit AdamW optimizer in the training process.

\section{Experiments and Case Studies}\label{sec-IV}
Subsection A presents comprehensive unit tests for the API-function matching module and the argument-value extraction module, specifically designed in Sections \ref{subsec-III(C)} and \ref{subsec-III(D)}, to validate their functionality. Subsection B revisits the warehouse problem introduced for LLMind 1.0 in Section \ref{sec-II}, but this time solves it using LLMind 2.0, demonstrating the improvements achieved with the distributed system. Subsection C introduces an additional case study in Wi-Fi networking, leveraging our real-world LLMind 2.0 testbed to validate the system’s practicality and feasibility.

General Experimental Setups: As in LLMind 1.0, we realize the coordinator with GPT4 to ensure a fair comparison, applying the same GPT4 parameter settings described in Section \ref{sec-II}: temperature = 0.5, top-p = 1.0, frequency penalty = 0.0. Note from Section \ref{sec-III} that the code generation pipeline within the agent requires two lightweight LLMs – one for the input-argument value extraction task in Step 2 and another for the final code generation in Step 3. In our agent implementation, we use Qwen3-0.6B for both models, with the model for Step 2 specifically fine-tuned according to the post-training process described in Section \ref{subsec-III(D)}.

\subsection{Unit Tests}
We first evaluate the subtask-to-API matching module in Step 1. This evaluation leverages the GPT4-generated test data obtained in Section \ref{subsec-III(D)} (recall that we reserved 10\% of the GPT4-generated data -- 120 devices with 4,800 API-function instances -- for testing). The aim of the experiment is to test the matching module’s accuracy in correctly selecting the desired API function under different sizes of the API RAG library, i.e., the number of API functions available in the selected device (henceforth denoted by $K$ for easier description). \footnote{Here we assume that the subtask has been correctly assigned to the device so that one of the $K$ APIs can match the subtask’s requirement – correctly assigning a subtask to the capable device belongs to the duty of the coordinator, and it has been investigated in LLMind 1.0, the predecessor of the current system developed.}

The evaluation process is  as follows: 

\begin{enumerate}
    \item \textbf{Sample Selection:} For each device in the test dataset, we randomly select $K$ samples of API functions from the device’s API library. As described in Section \ref{subsec-III(D)}, each sample consists of a JSON-formatted brief of the API function and a subtask description that matches the API function. Since each device in the testing dataset has 40 APIs, we must ensure $K \leq 40$ in the testing process. 

    \item \textbf{RAG Embedding Construction:} For each of the $K$ selected test samples, we use the API-function description within the JSON-formatted API function brief to construct the RAG embedding for that sample (see Fig. \ref{fig:9} for details).
 
    \item \textbf{Subtask Matching for Each Device:} We randomly select a subtask description from the $K$ samples. Specifically, with reference to Fig. \ref{fig:A1}, this subtask description is converted to a vector embedding. Meanwhile, we also have the $K$ vector embeddings of $K$ API functions (note: each vector embedding is produced from the “notes” part of an API function description in the JSON format). We identify the API-function embedding most similar to the subtask-description embedding and determine that as a match.  

    \item \textbf{Repeat and Averaging:} For each device, we repeat steps 1) to 3) 100 times. In each iteration, we randomly select $K$ APIs out of the 40 available APIs from step 1), and in step 3), we randomly select one API out of the $K$ APIs.  After calculating the average accuracy of the matching module on that device across the 100 iterations, we proceed to the next device. Once all 120 devices in the testing dataset have been evaluated, we calculate the overall average accuracy to determine the matching module's performance across the entire dataset.
\end{enumerate}

\begin{figure}[htbp]
  \centering
  \includegraphics[width=0.475\textwidth]{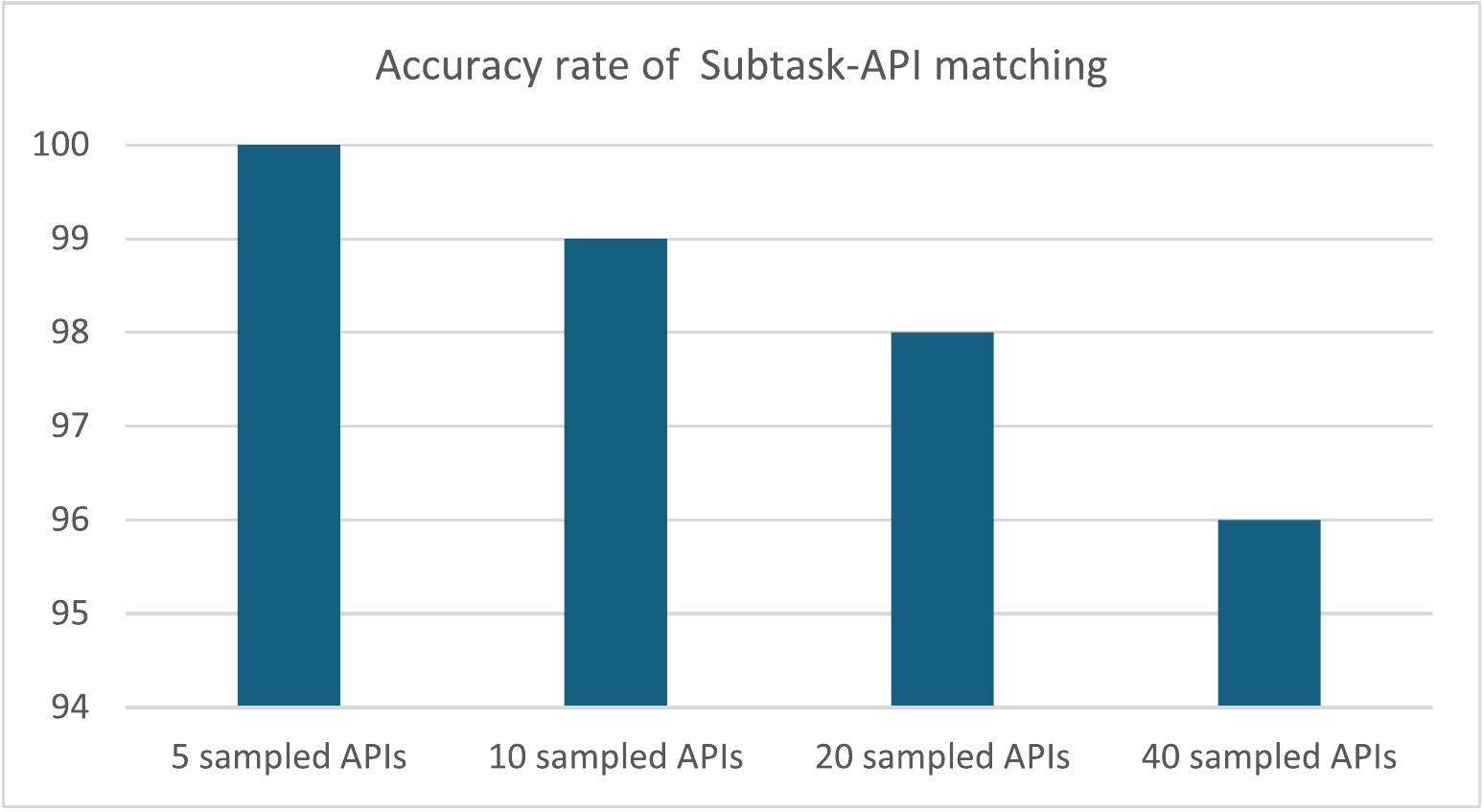}\\
  \captionsetup{font={small}}
  \caption{Accuracy statistics showcasing the reliability of the subtask-to-API matching module for different sizes of RAG library (i.e., the value of $K$) are considered for IoT devices.}
\label{fig:10}
\end{figure}

We consider four possible $K$ values (i.e., 5, 10, 20, and 40) in our experiments and present the result in Fig. \ref{fig:10}. We see from the figure that the matching module maintains highly reliable for cases with $K=5$ and $K=10$. While the module’s accuracy decreases when $K$ reaches 20, we highlight that performance remains reliable – even for extreme cases with $K=40$, the accuracy is still larger than 90\%. Importantly, we note that a practical IoT device may have a very limited number of APIs - most of them have less than 10, or even no more than 5 APIs. Based on this observation, we are confident in the performance of the matching models in practical IoT applications.

We next examine the performance of the argument-value extraction module in Step 2 of the code generation pipeline, using the test dataset obtained in Section \ref{subsec-III(D)}. We also test the following model for comparison: 1) powerful cloud-based models, such as Gemini 2.5 Pro, Chaude-3.7, and GPT4, and 2) Qwen-3 models of different sizes, including the lightweight 0.6B model that serves as our foundation model for the fine-tuning process.

\begin{figure}[htbp]
  \centering
  \includegraphics[width=0.475\textwidth]{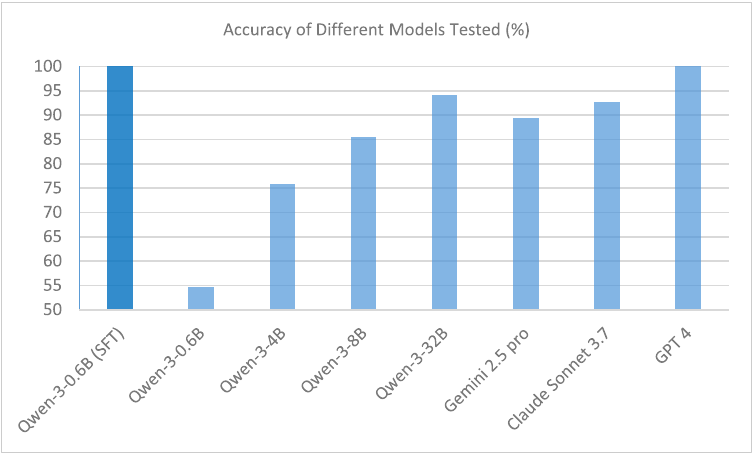}\\
  \captionsetup{font={small}}
  \caption{Accuracy statistics showcasing the ability of the different LLMs on the input-argument value extraction task.}
\label{fig:11}
\end{figure}

Fig. \ref{fig:11} shows the accuracy of these models on the test dataset. While advanced cloud-based models have left us with an impression of being powerful for general tasks, only GPT4 could guarantee 100\% reliability in our task. As for the non-fine-tuned Qwen-3 series, the non-fine-tuned Qwen 0.6B has the worst performance, with accuracy of only 54.77\%. However, after fine-tuning, it also achieves 100\% accuracy. This observation highlights the significant performance boost through fine-tuning.

For the above experiment, we would like to clarify two subtle technical details. First, regarding the evaluation standard, our experiment considers a test sample successful only if all API argument values are correctly extracted from the human command in the correct order (recall from Section \ref{sec-III} that the data generation process includes APIs with multiple arguments.) Second, the experimental results in Fig. \ref{fig:11} are intended to highlight the effectiveness of model fine-tuning. The model’s generalization ability --- specifically its performance in handling practical APIs and human commands --- is discussed in Fig. \ref{subfig13.c}. Readers interested in further details could refer to the discussion therein.

\subsection{Case Study One - The Warehouse Vacancy Search Task}
Recall from Section \ref{sec-II} that we have a warehouse case study, which considers a fixed number of shelves and a variable number of mobile robots of various brands. With this setup, a human warehouse manager gives the following instruction to the system: \textit{"Please check if there are vacant positions on the shelves"}. After receiving the command, the coordinator assigns the vacancy search task to distributed robots (see Section \ref{sec-II} for detailed description of the task). 

Fig. \ref{fig:3} in Section \ref{sec-II} presents the processing latency and the success rate of code generation when LLMind 1.0 is used. This subsection considers what if LLMind 2.0 is used instead, following the same experimental setup as in Fig. \ref{fig:3}.

\begin{figure}[htbp]
  \centering
  \includegraphics[width=0.49\textwidth]{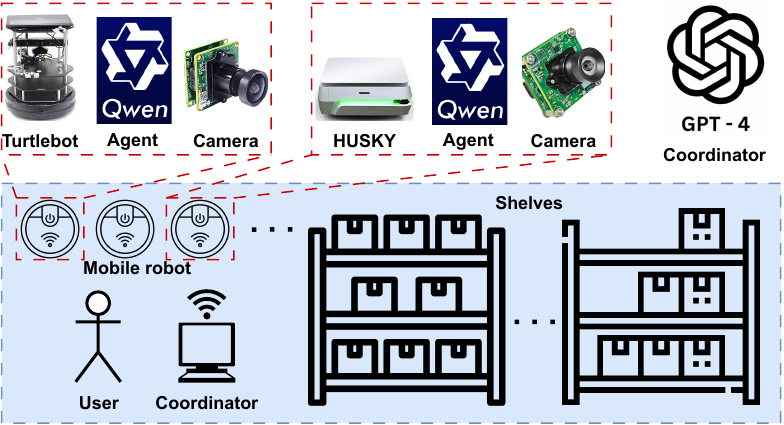}\\
  \captionsetup{font={small}}
  \caption{An illustration of the tested warehouse scenario when LLMind 2.0 is considered.}
\label{fig:12}
\end{figure}

\begin{figure}[htbp]
    \centering
    \subfloat[]
    {
      \label{subfig13.a}
      \includegraphics[width=0.425\textwidth]{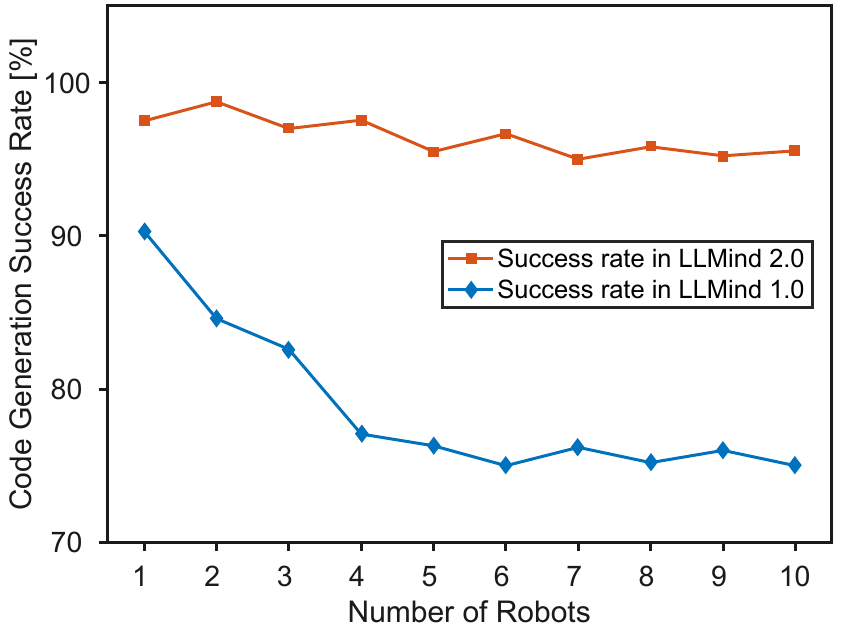}
    }%
    \hfill
    \subfloat[]
    {
      \label{subfig13.b}
      \includegraphics[width=0.425\textwidth]{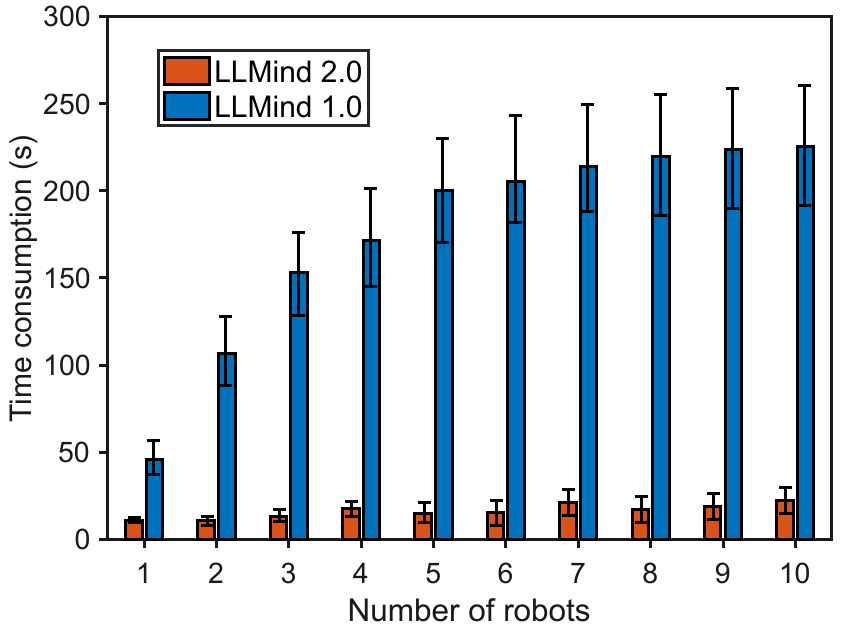}
    }
    \\
    \subfloat[]
    {
      \label{subfig13.c}
      \includegraphics[width=0.45\textwidth]{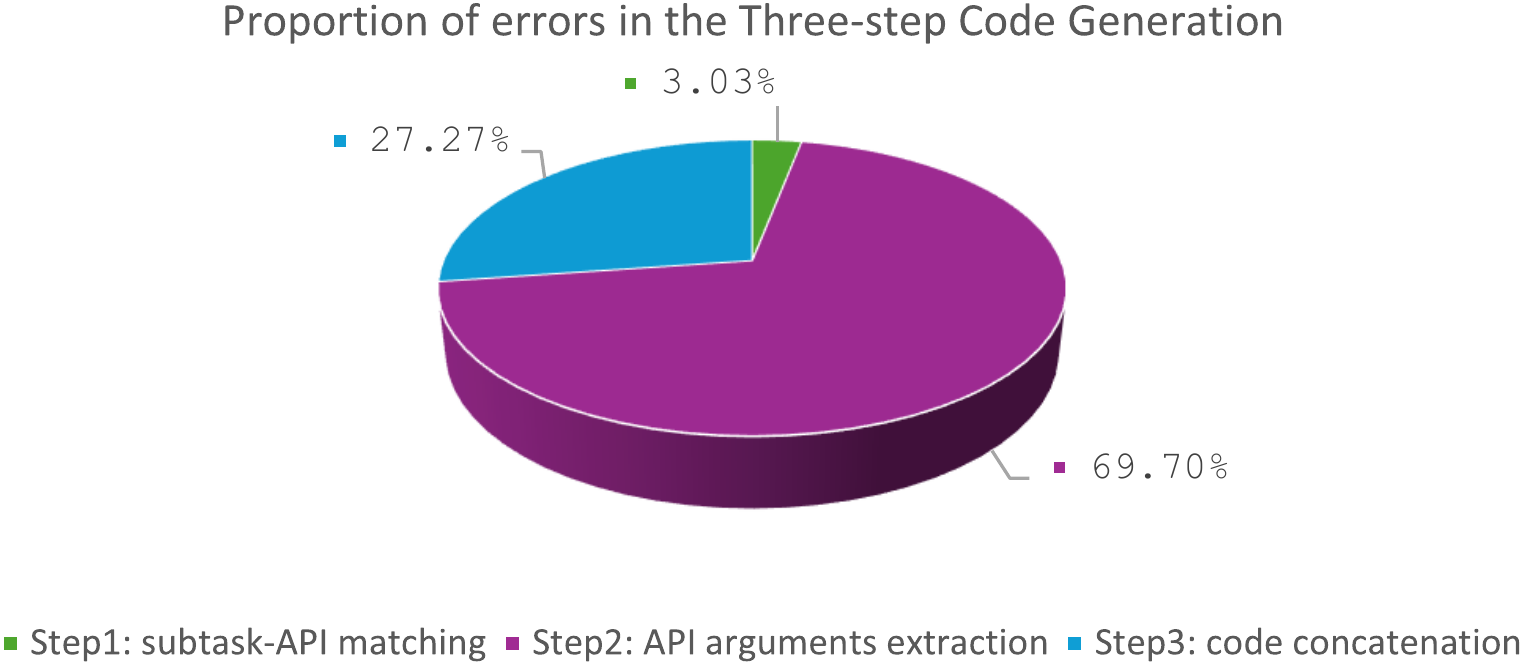}
    }
    \caption{Experimental results showcasing LLMind 2.0’s improved success rate in code generation.}
    \label{fig:13}
\end{figure}

Fig. \ref{fig:12} illustrates the system setup in this case study when LLMind 2.0 is considered. Experimental results are presented in Fig. \ref{fig:13}. From Fig. \ref{subfig13.a}, we see LLMind 2.0’s enhanced reliability compared to LLMind 1.0. For LLMind 1.0, since the task is solely undertaken by the central coordinator and can be complex when a larger number of robots are involved, the rate of successful code generation decreases when more robots are considered. Under the LLMind 2.0 setup, on the other hand, the subtask assigned to a distributed agent remains simple regardless of the number of robots, as the agent takes care of one robot only. Therefore, regardless of the number of robots, LLMind 2.0 always shows a very reliable performance in terms of successful code generation for all robots. Meanwhile, as shown in Fig. \ref{subfig13.b}, significant reductions in time consumption are also achieved thanks to parallel code generations.

Fig. \ref{subfig13.c} takes a close look at the causes of LLMind 2.0’s code generation failures. Following the experimental setup in Section \ref{sec-II}, each data point in Fig. \ref{subfig13.a} comes from 100 repeated experiments. Therefore, for the orange line representing LLMind 2.0 in Fig. \ref{subfig13.a}, we have 1000 independent tests in total and 33 of them ended with failures. Among these failures, only 1 case is caused by mistaken subtask-API mapping in Step 1, and most errors happen in Steps 2 and 3.

Step 2 accounts for 69.70\% of the errors (i.e., 33 out of 1000 tests failed, with 23 attributed to Step 2). In contrast to the 100\% accuracy achieved by the fine-tuned 0.6B model in the unit test (see Fig. \ref{fig:11}), the Step 2 errors shown in Fig. \ref{subfig13.c} are likely caused by the fine-tuned model’s limited generalization ability when trained exclusively on artificially generated data via GPT4’s role-playing. Specifically, while the fine-tuned model achieves 100\% accuracy on GPT4-generated data, practical robot APIs may differ from AI-generated APIs. This discrepancy can lead to argument-value extraction errors when practical scenarios deviate significantly from the training scenarios.

This issue is a common challenge in model distillation based on LLM-generated answers \cite{hsieh2023distilling, ouyang2022training}. A potential solution involves incorporating a large volume of real-world API and subtask description data, extracted from working IoT devices, into the training dataset. However, this approach requires substantial manpower and experimental effort to build the dataset. Additional strategies for addressing the generalization problem are discussed at the end of Section \ref{sec-V}, and we plan to explore these solutions as part of our future work.

\begin{figure}[htbp]
  \centering
  \includegraphics[width=0.475\textwidth]{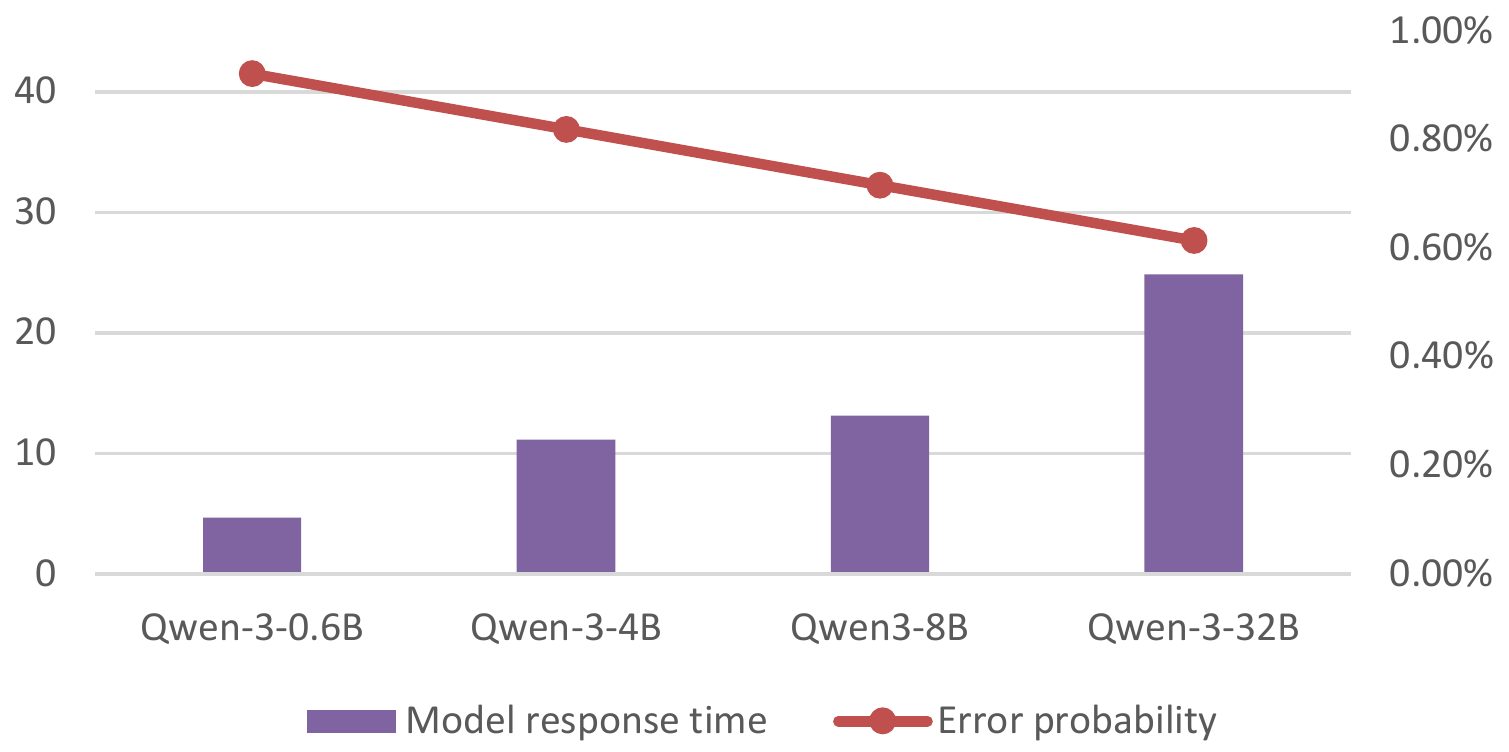}\\
  \captionsetup{font={small}}
  \caption{Error probabilities and response times of Qwen3 models with different sizes when given the task in Step 3.}
\label{fig:19}
\end{figure}

Step 3 accounts for 27.27\% of the errors (i.e., 33 out of 1000 tests failed, with 9 of these failures attributed to Step 3). This is likely due to the limitations of the Qwen3-0.6B pre-trained model in code generation. This hypothesis is supported by the experiment shown in Fig. \ref{fig:19}, where we evaluated the error probability of larger-scale Qwen3 models on the same task. The results suggest that errors in Step 3 can be mitigated by replacing the current model with more powerful pre-trained models, albeit at the cost of increased response time.

We anticipate that this issue can be resolved through future advancements in lightweight pre-trained models. Next-generation lightweight LLMs are expected to deliver improved coding performance without increasing model size or response time.

\subsection{Case Study Two: System Demo in WiFi Network}
This subsection deploys an LLMind 2.0 system in a WiFi system to validate the feasibility of the proposed system. We develop two experiments to showcase how LLMind 2.0’s operating protocol facilitates human-like competition and collaboration behaviors among distributed devices with device agents’ information report sessions and the coordinator’s decision making with global considerations. Meanwhile, these experiments also highlight how an agent composes control scripts tailored to the associated device to realize advanced functions of the device beyond the standard WiFi networking protocol, and how the coordinator mentors such operations to ensure the overall system is operated in a fair and graceful manner.

Let us start with a quick \textbf{system setup overview}.

\begin{figure}[htbp]
  \centering
  \includegraphics[width=0.49\textwidth]{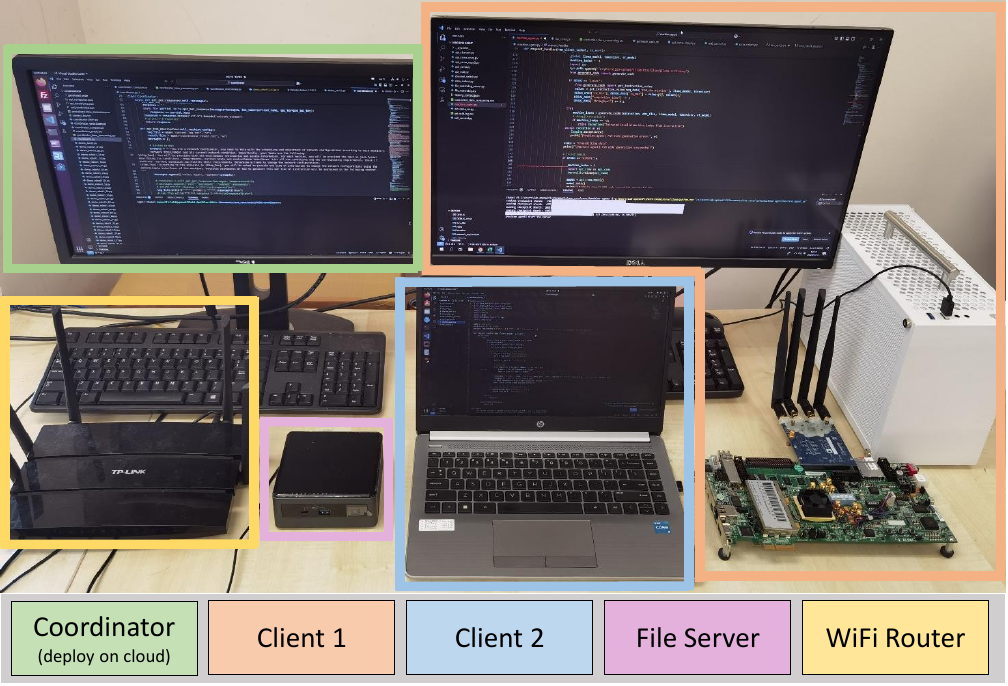}\\
  \captionsetup{font={small}}
  \caption{A real-world IoT system developed to showcase the collaborative behavior of LLMind 2.0 device agent.}
\label{fig:14}
\end{figure}

\textit{General System Setup:} Fig. \ref{fig:14} illustrates the system setup for experiments in this subsection. Two WiFi clients connect to a WiFi access point (AP) with the IEEE 802.11n standard. Client 1 uses an FPGA-based software-defined radio (SDR) platform that runs OpenWiFi to realize WiFi functionality; Client 2 has a commercial WiFi adapter. Both clients support dual-band WiFi functionality, and the AP can support 2.4 GHz and 5 GHz WiFi connections simultaneously. The AP has a wired connection to a mini-PC serving as the file server. The two clients keep uploading high-resolution images via TCP to the file server, with each image file having a 4MB fixed size. Both the clients and the server work on the Ubuntu 18.04 operating system.

\textit{LLMind 2.0 Deployment:} To run the LLMind 2.0 system, the two clients are equipped with local device agents, with each agent supported by an GTX 1060 GPU – a representative setup in terms of the computational ability of IoT devices. 

\textit{API Functions:} The wireless network inference cards (NICs) in Client \#1 and Client \#2 are realized with the OpenWifi project running on an FPGA chip and the RTL8812BU chip, respectively. Both the SDR project and the commercial chip have open-sourced their Linux driver on GitHub, available at \cite{openwifi} and \cite{RTL8812BU}, respectively. Detailed introduction about the API functions supported by each NIC is available at the project’s GitHub description.

\begin{figure}[htbp]
  \centering
  \includegraphics[width=0.49\textwidth]{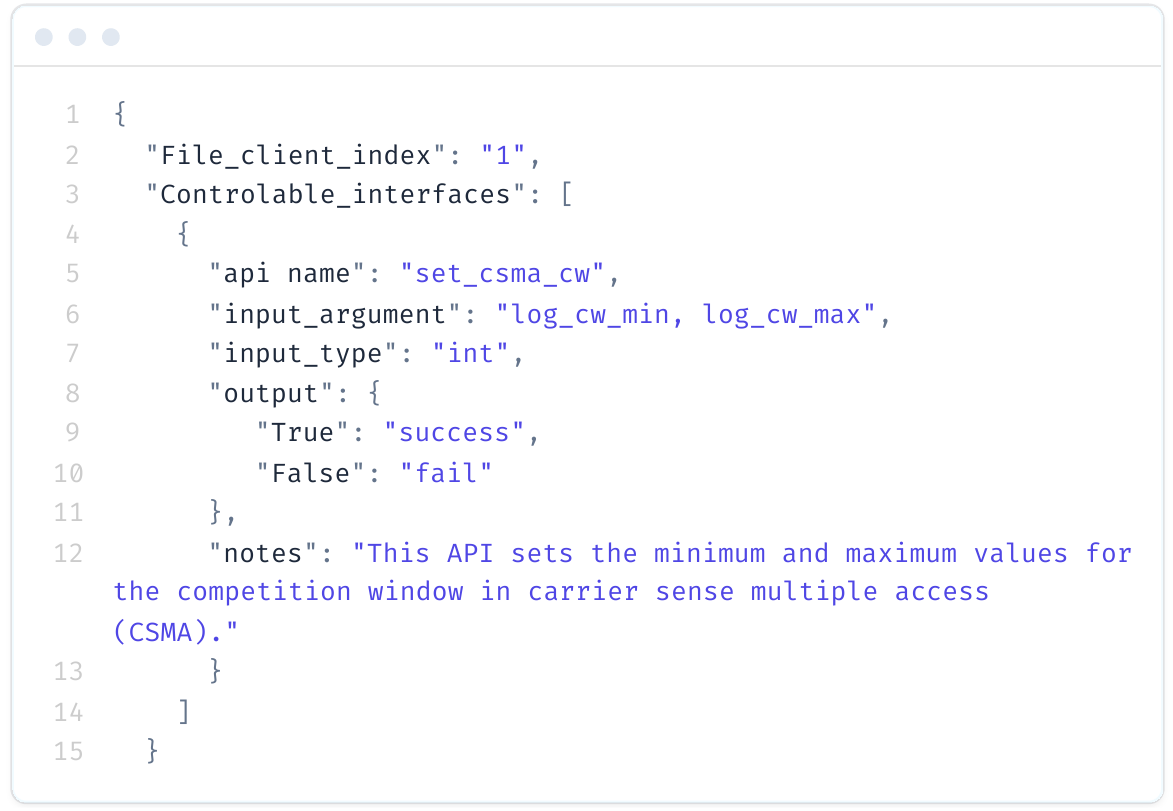}\\
  \captionsetup{font={small}}
  \caption{CW\_min and CW\_max configuration API for OpenWiFi. Note that the input arguments of the API function are the log of the CW\_min and CW\_max.}
\label{fig:15}
\end{figure}

The OpenWiFi-based NIC in Client \#1 offers more flexible operations compared with standard commercial NICs. For WiFi, transmission collisions necessitate a backoff operation. The NIC selects a random number between CW\_min and CW\_max and uses this number as the length of its contention window (CW) \cite{chang2005ieee}. In commercial NICs, the values of CW\_min and CW\_max are fixed to predefined settings. However, the OpenWiFi-based NIC allows for manual configuration of CW\_min and CW\_max, although the standard values are adopted by default. The flexibility to overwrite the default values makes the control of the NIC more customizable. Fig. \ref{fig:15} illustrates the JSON-formatted API function description provided by OpenWiFi for configuring CW\_min and CW\_max \cite{jiao2020openwifi}.

On the other hand, the NIC built upon SDR platforms has its disadvantages: its reliability is not as good as a commercial chip under the same channel condition. This is because the SDR project, developed for demonstration purposes, has neither realized advanced coding/decoding algorithms nor gone through comprehensive testing as a commercial chip may have done for optimized performance. And it has also been observed that the SDR-based NIC, compared with the commercial NIC, tends to exhibit lower signal transmission power.\footnote{According to our analysis of open discussions within the OpenWiFi community, the lower signal transmission is probably due to the absence of power amplifiers in the AD9361 RF extension board.} These reasons lead to the Client 1 NIC’s lower signal-to-noise ratio (SNR) and higher packet-error rate (PER) when compared with the commercial NIC used in Client 2.

With the above background, we now introduce the experiment conducted in \textbf{Scenario 1}. 

This experiment considers an image uploading task, in which each device keeps uploading images to the file server in a one-after-another manner. We want the successful uploading of each 4MB image to be accomplished within a pre-defined time threshold (here we set the timing requirement as 16 seconds). At the beginning of the experiment, Client 1 was the only device within the network, exclusively occupying all network resources to achieve the desired performance. At that time, the OpenWiFi-based NIC in Client 1 has a default \textit{log\_CW\_min} and \textit{log\_CW\_max} setup of 10 and 15, respectively. We denote this setup by (10,15).

\begin{figure}[htbp]
  \centering
  \includegraphics[width=0.49\textwidth]{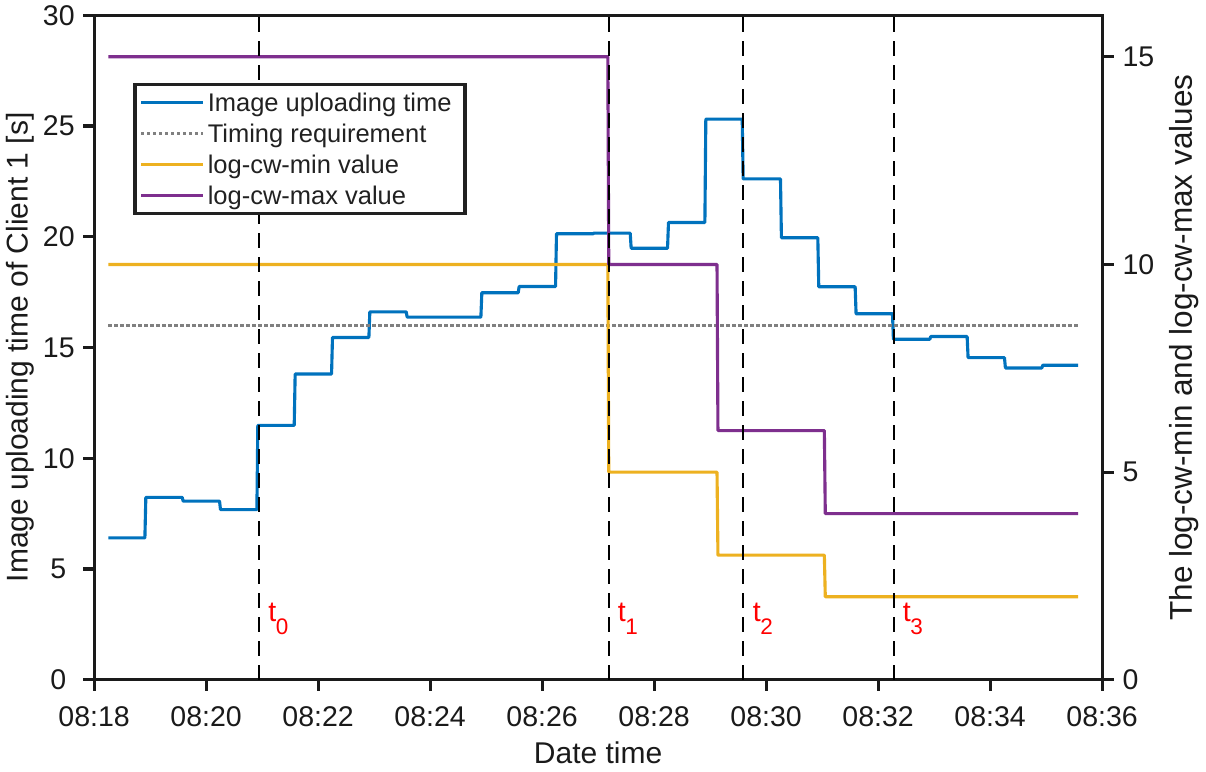}\\
  \captionsetup{font={small}}
  \caption{Experimental observation in Scenario 1. Note from this figure that there is a lag effect in the CW value adjustment of Client 1, which is attributed to the Cubic congestion control algorithm deployed in the Linux TCP stack. The lagging effect found in the experimental result is consistent with the observation reported in \cite{ha2008cubic}.}
\label{fig:16}
\end{figure}

Shortly thereafter ($t_0$ in Fig. \ref{fig:16}), Client 2 joined the network and began image uploading. The two clients competed for transmission opportunities, with each client obtained around 50\% of the network resources - the CSMA protocol ensured fair transmissions for all devices within the network when they had the same default setup for CW\_min and CW\_max. However, as the OpenWiFi-based NIC in Client 1 had higher probability of packet-transmission failure (i.e., more packet retransmissions under the TCP protocol), it required more than half of transmission opportunities to fulfill the 16-second timing requirement. Therefore, as seen in Fig. \ref{fig:16} (from $t_0$ to $t_1$), Client 1 had problems meeting the timing requirement after Client 2 joined the network. On the other hand, Client 2, (the timing data is not presented in Fig. \ref{fig:16})  had a WiFi connection with a more reliable commercial chip and could meet the 16-second timing requirement easily.

The report of LLMind 2.0 device agents (see Fig. \ref{fig:8}) and the coordinator’s decision-making process after collecting the agent reports (see Fig. \ref{fig:7}) enable coordinated resource sharing. After receiving the periodic polling message from the coordinator, a device agent reports its state, including the device’s latest MAC and PHY configurations (such as the setup of CW\_min and CW\_max) and its application layer information (such as its task fulfillment result, i.e., whether the device has uploaded the file in a timely manner), to the coordinator. After receiving each device’s state report, the coordinator, which is realized with a powerful cloud-based LLM with strong reasoning ability, analyzes the current system’s performance from a global perspective and makes adjustments for these devices as needed. In this specific experiment, after noticing that Client 2 could easily fulfill the uploading task while Client 1 had problems making it, the coordinator made the following decisions: increase the proportion of network resources allocated to Client 1 while ensuring Client 2 still has the necessary resources to fulfill its task.

As can be seen from the experiment record (video available at \textcolor{blue}{\url{https://youtu.be/kxFzIE9v9Ik}}, the coordinator asked Client 1 to reduce its \textit{log\_CW\_min} and \textit{log\_CW\_max} values for more aggressive competition for the airtime, helping the SDR-driven NIC to obtain more than half of the transmission opportunities.\footnote{This improvement could be attributed to the EDCF mechanism in WiFi networks \cite{chang2005ieee}, where lowering the CW value can reduce Client 1's backoff time in CSMA, thus giving more packet transmission opportunity to Client 1.}  Importantly, the adjustment process supervised by the coordinator happened in a gradual and incremental manner to ensure the task fulfillment of Client 2 was not affected. As we can see from Fig. \ref{fig:16}, the initial setup of (\textit{log\_CW\_min}, \textit{log\_CW\_max}) for Client 1 was (10, 15) at $t_1$, then the configuration dropped to (10, 5), (5, 10) and (3, 6) following the coordinator’s advice. After these adjustments, as the coordinator found that 1) Client 1 still had problems meeting the 16-second timing requirement, and 2) Client 2 could still meet the timing, it asked Client 1 to further reduce the configuration to (2, 4). Finally, the image uploading time of Client 1 decreased to 15.37 seconds at time $t_3$, falling below the timing requirement for file uploading (and Client 2 still has no problem meeting that timing requirement). This process can also be viewed as a benign competition of networking resources between agent 1 and agent 2 under the supervision of the coordinator. 

The above experiment shows how LLMind 2.0’s key features, i.e., state reports from distributed agents and the coordinator’s global decision making, enable dynamic system adjustment in a fair and graceful manner. We emphasize that using LLM-empowered agents for device control allows the device to better utilize its hardware via API controls. In this case, the device agent leverages the flexibility of the SDR-based NIC in Client 1 by realizing dynamic controls of the CSMA contention window, which is a special feature not supported by IEEE 802.11 standard. The dynamic contention window control, under the supervision of the coordinator, results in more reasonable resource allocations that help both devices to meet the application’s timing requirements, as opposed to pure MAC-layer competition in the standard WiFi set-up.

\textbf{Scenario 2} considers another scenario where the two devices tried to complete the image transmission task under strong interference. The interference was caused by a microwave oven in the environment (very close to those devices shown in Fig. \ref{fig:14}, but not shown in the image). Unlike the timing requirement in the previous experiment, the requirement for both clients here is to \textit{\textbf{maintain packet-error rate (PER) not exceeding 20\%}}. At the beginning of the experiment, both Clients 1 and 2 were connected to the WiFi AP via a common 2.4 GHz channel and they kept uploading 4M images via UDP to the server. After some time, a microwave oven was turned on (microwave ovens typically operate at a frequency band around 2.4 GHz). The following experiment shows how the microwave oven’s inference would affect the PER of both clients and how the two clients collaborated through information sharing to overcome the strong inference by changing to the 5GHz channel.

\begin{figure}[htbp]
  \centering
  \includegraphics[width=0.49\textwidth]{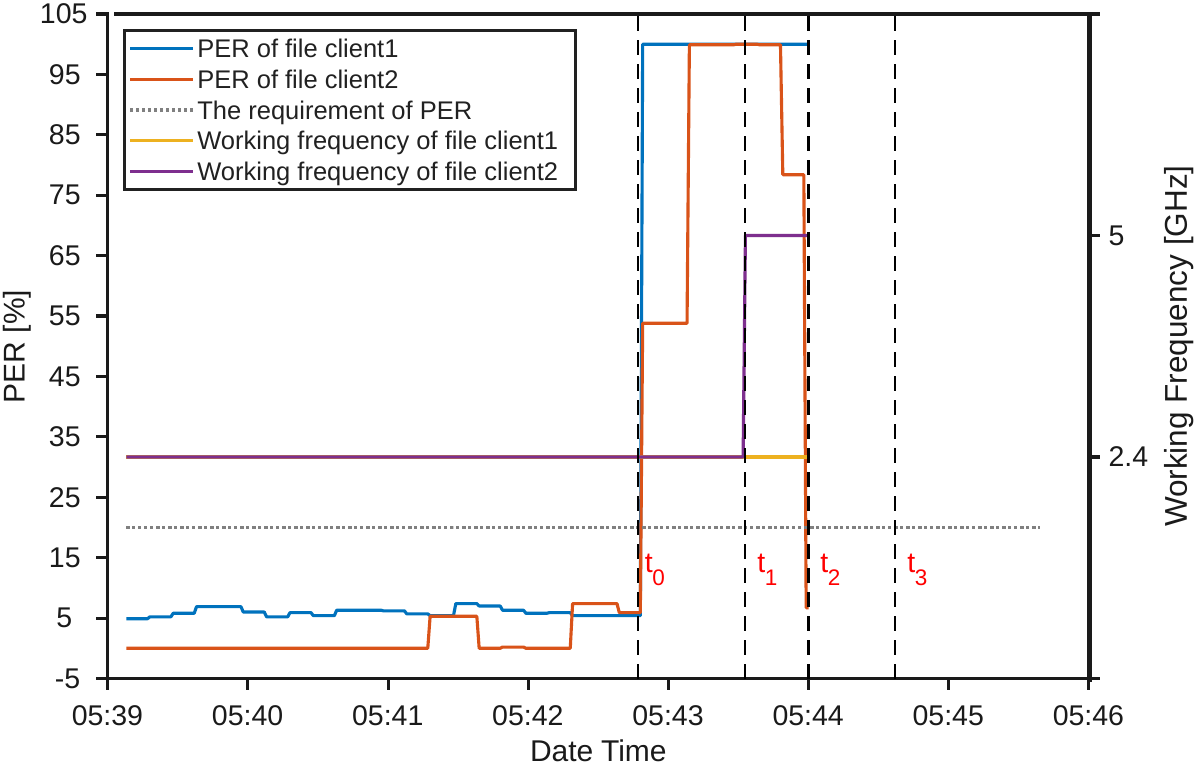}\\
  \captionsetup{font={small}}
  \caption{Experimental observation in Scenario 2, in which the UDP protocol is applied. Due to the limitation of SDR, Client 1 cannot switch from the 2.4 GHz connection to the 5 GHz connection without rebooting. Client 2, on the other hand, uses a commercial WiFi NIC that allows channel switching without system rebooting.}
\label{fig:17}
\end{figure}

Fig. \ref{fig:17} presents the two clients’ PER during the experiment. Before $t_0$, the PER values of both clients met the expected performance requirements (i.e., less than 20\% PER). When the microwave oven was turned on at $t_0$, both devices suffered from high PER due to the interference.

An advantage of Client 1 is that the SDR-based NIC therein allows the agent to collect more detailed lower-layer network information than a commercial NIC. That is, an SDR NIC provides the agent with detailed information about the process within PHY in MAC, while a commercial NIC may only provide the agent with limited information allowed by the chip manufacturer. Therefore, when faced with the strong 2.4GHz inference, the agent in Client 1 could provide detailed channel sensing information collected by the OpenWiFi project in its state report to the coordinator. Thanks to this information, the coordinator becomes aware of the wireless inference and informed Client 2 about the interference and asked it to switch to the AP’s 5 GHz channel. Following a disconnection of the 2.4GHz link and a reconnection to the 5GHz link, Client 2's PER value drops to 6.7\% at time $t_2$, meeting the required PER performance. On the other hand, Client 1 was unable to adjust its channel selection because the SDR platform does not allow channel switching without system reboot. Its PER went back to normal when we turned off the microwave at $t_3$. A detailed video demonstration is available at \textcolor{blue}{\url{https://youtu.be/zAfkCm1sRHQ}}. 

In the above experiment, Client 1 shared its sensing information in a selfless manner, i.e., helping Client 2 out of trouble even if itself could not get rid of the inference on the 2.4GHz channel. Unlike the resource competition in the previous experiment, this experiment shows that the LLMind 2.0 operation framework also allows cooperative behaviors between device agents, which is realized by agent reports to the coordinator and the coordinator’s decision-making with global consideration.

\section{Conclusion and Outlooks}\label{sec-V}
This paper presents LLMind 2.0, a distributed IoT automation framework designed to address the challenges of managing large-scale IoT devices using embedded LLM-powered device agents and natural language-based machine-to-machine communication. Unlike traditional centralized approaches, LLMind 2.0 delegates device-specific code generation tasks to device agents equipped with fine-tuned LLMs tailored for control script generation. This distributed architecture reduces the computational burden on the central coordinator and enables seamless integration and collaboration across heterogeneous devices, regardless of their manufacturer or supported programming language.

The framework incorporates several key technical components, including a retrieval-augmented generation (RAG) mechanism for accurate subtask-to-API function mapping and a finite state machine-based code generation pipeline. Combined with task-specific fine-tuned models, these components enhance both the reliability and efficiency of the automation system. Furthermore, by localizing sensitive code generation processes to device agents, LLMind 2.0 ensures that proprietary device data remains on-device, providing significant privacy and security benefits.

Experimental evaluations conducted in multi-robot warehouse scenarios demonstrate that LLMind 2.0 outperforms centralized baselines in terms of scalability, latency, and task execution success rates. Additional experiments in practical WiFi deployments reveal that LLMind 2.0’s unique collaborative decision-making mechanism—realized through device agents’ state reports and coordinator-driven global decision-making—enables effective collaboration and resource-efficient competition among distributed devices.

Overall, our results suggest that natural language can serve as a practical and effective medium for both human-to-machine and machine-to-machine interactions in IoT systems, enabling greater flexibility, parallelism, and device collaboration. By open-sourcing the code and fine-tuning datasets, we aim to foster further research into distributed, language-driven automation systems. 

Future work could focus on scaling LLMind 2.0 to support larger device ecosystems and enabling more sophisticated collaborative behaviors to further enhance the autonomy and practicality of IoT environments. Additionally, future efforts may aim to improve the generalization ability of the fine-tuned model by incorporating real-world API and subtask description data extracted from operational IoT devices. Another potential technical approach for enhancing model generalization could involve the recently proposed co-evolutionary reinforcement learning (RL) framework, which has demonstrated effectiveness in generating high-quality, LLM-synthesized data for building self-evolving LLM agents

\appendices
\section{\text{ } \text{ } \text{ } Robot API}\label{sec-App1}
The API of a mobile robot encompasses both the robot's intrinsic functionalities and those provided by its peripheral modules, such as the wireless network module, camera, and additional custom functions tailored for the warehouse scenario. Specifically, the mobile robot offers a comprehensive set of intrinsic API functions, including retrieving battery status, initiating automatic docking, performing emergency stops, adjusting movement speed, creating environmental maps, localizing itself, and navigating based on positional coordinates.

In addition to these intrinsic functions, the camera module provides API functions for adjusting the shooting angle and capturing photos, while the wireless network module offers functions to retrieve available WiFi hotspots, connect to networks, and enable AP mode. Moreover, the robot includes custom API functions specific to the warehouse scenario, such as navigating to designated shelves and identifying vacancies or items on those shelves.

\begin{figure*}[htbp]
  \centering
  \includegraphics[width=\textwidth]{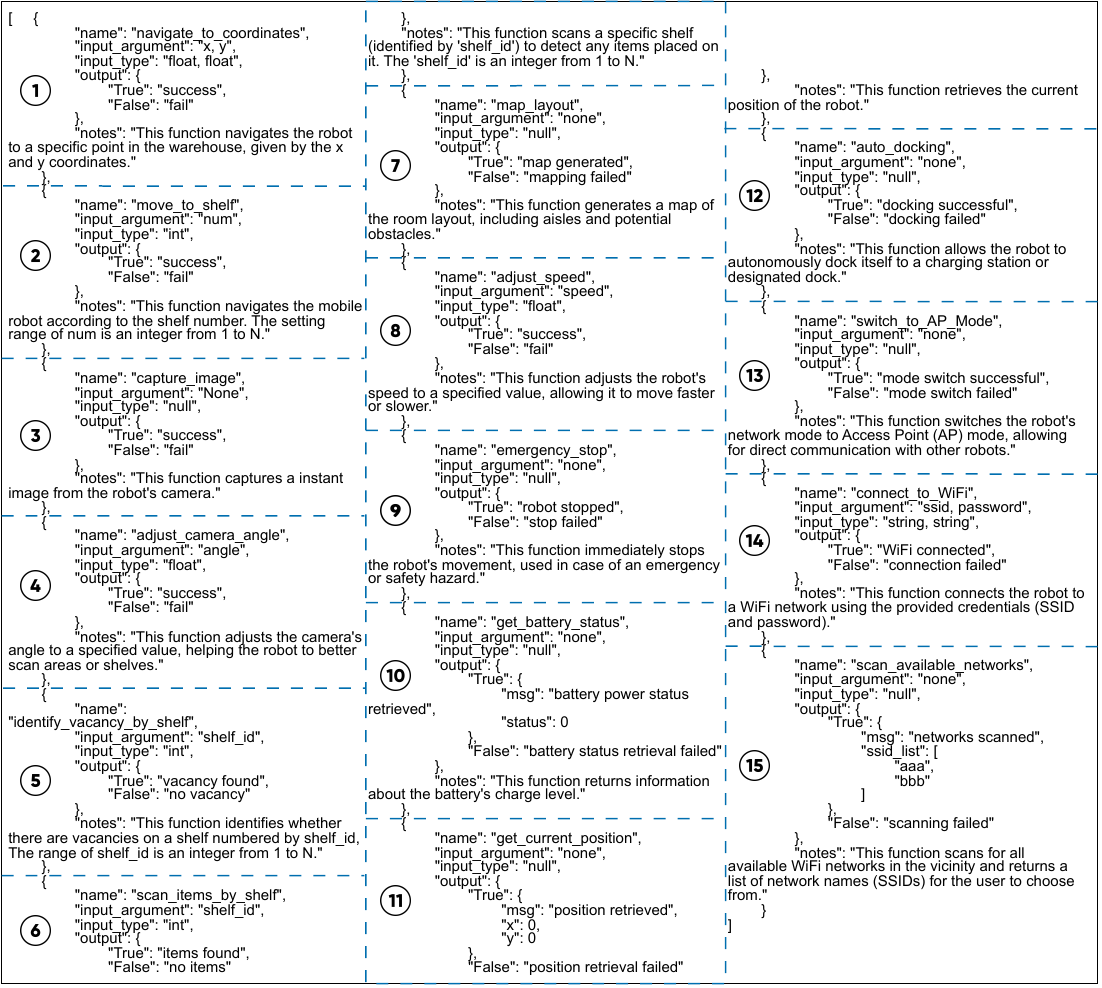}\\
  \captionsetup{font={small}}
  \caption{High-level JSON-formatted robot API descriptions.}
\label{fig:A1}
\end{figure*}

Fig. \ref{fig:A1} presents high-level descriptions of these APIs, detailing essential information for each function in a structured JSON format. This includes the function name, input arguments with associated data types, return outputs, and a concise description of the function's purpose. These high-level API descriptions are provided to the coordinator to enhance its understanding of the robot's capabilities, serving as a reference for the coordinator to effectively allocate and arrange subtasks.

\begin{figure}[!ht]
  \centering
  \includegraphics[width=0.5\textwidth]{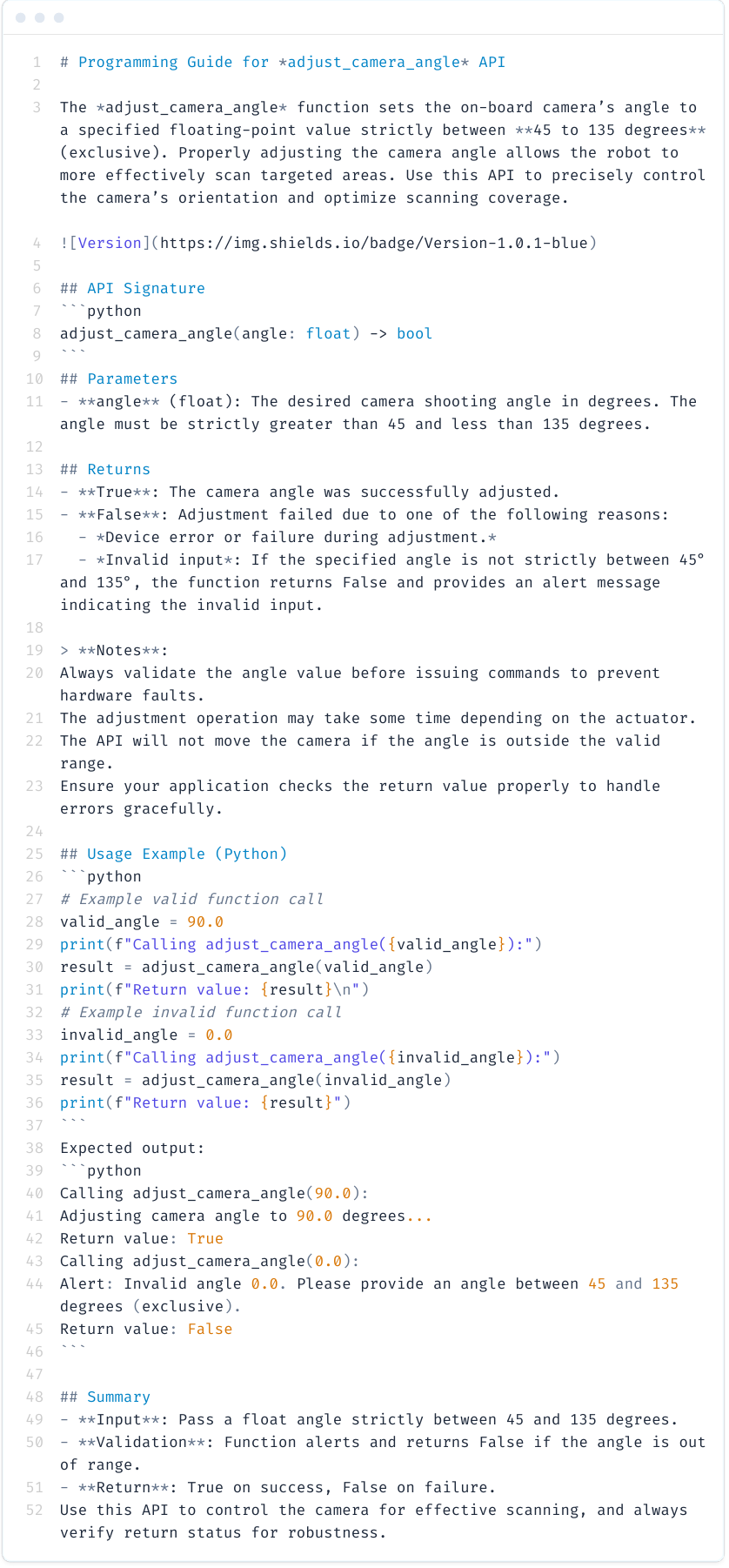}\\
  \captionsetup{font={small}}
  \caption{A sample of the content in the detailed user guides: “adjust\_camera\_angle” in  Fig. \ref{fig:A1} for example.}
\label{fig:A2}
\end{figure}

In comparison to high-level descriptions in Fig. \ref{fig:A1}, Fig. \ref{fig:A2} gives an example of detailed API programming guidance available in a device handbook, where design details of the API call are elaborated. Detailed device API user guides are not provided to the coordinator, given that 1) the coordinator does not need such details for subtask design, and 2) long contexts for each API in each device can overwhelm the coordinator LLM. Instead, the user guide is available in the associated IoT devices so that the device agent can leverage the guidance for script generation.

\bibliographystyle{IEEEtran}
\bibliography{Main}

@article{deng2025exploring,
  title={Exploring {DeepSeek}: A Survey on Advances, Applications, Challenges and Future Directions},
  author={Deng, Zehang and Ma, Wanlun and Han, Qing-Long and Zhou, Wei and Zhu, Xiaogang and Wen, Sheng and Xiang, Yang},
  journal={IEEE/CAA Journal of Automatica Sinica},
  volume={12},
  number={5},
  pages={872--893},
  year={2025},
  publisher={IEEE}
}

@article{cui2024llmind,
  title={{LLMind}: Orchestrating {AI} and {IoT} with {LLM} for complex task execution},
  author={Cui, Hongwei and Du, Yuyang and Yang, Qun and Shao, Yulin and Liew, Soung Chang},
  journal={IEEE Communications Magazine},
  year={2024},
  publisher={IEEE}
}

@article{xiao2024efficient,
  title={Efficient prompting for {LLM-based} generative {Internet of Things}},
  author={Xiao, Bin and Kantarci, Burak and Kang, Jiawen and Niyato, Dusit and Guizani, Mohsen},
  journal={IEEE Internet of Things Journal},
  year={2024},
  publisher={IEEE}
}

@inproceedings{kok2024iot,
  title={When {IoT} meet {LLMs}: Applications and challenges},
  author={K{\"o}k, {\.I}brahim and Demirci, Orhan and {\"O}zdemir, Suat},
  booktitle={2024 IEEE International Conference on Big Data (BigData)},
  pages={7075--7084},
  year={2024},
  organization={IEEE}
}

@article{qin2023toolllm,
  title={{ToolLLM}: Facilitating large language models to master 16000+ real-world {APIs}},
  author={Qin, Yujia and Liang, Shihao and Ye, Yining and Zhu, Kunlun and Yan, Lan and Lu, Yaxi and Lin, Yankai and Cong, Xin and Tang, Xiangru and Qian, Bill and others},
  journal={arXiv preprint arXiv:2307.16789},
  year={2023}
}

@article{chen2024octopus,
  title={Octopus: On-device language model for function calling of software {APIs}},
  author={Chen, Wei and Li, Zhiyuan and Ma, Mingyuan},
  journal={arXiv preprint arXiv:2404.01549},
  year={2024}
}

@article{patil2023gorilla,
  title={Gorilla: Large language model connected with massive {APIs}, 2023},
  author={Patil, Shishir G and Zhang, Tianjun and Wang, Xin and Gonzalez, Joseph E},
  journal={URL https://arxiv.org/abs/2305.15334},
  year={2023}
}

@article{chang2005ieee,
  title={{IEEE} 802.11 {DCF}},
  author={Chang, Presented By Yu Chu},
  journal={IEEE Transactions on Mobile Computing},
  volume={4},
  number={4},
  year={2005}
}

@article{gao2023retrieval,
  title={Retrieval-augmented generation for large language models: A survey},
  author={Gao, Yunfan and Xiong, Yun and Gao, Xinyu and Jia, Kangxiang and Pan, Jinliu and Bi, Yuxi and Dai, Yixin and Sun, Jiawei and Wang, Haofen and Wang, Haofen},
  journal={arXiv preprint arXiv:2312.10997},
  volume={2},
  number={1},
  year={2023}
}

@electronic{SentenceTransformer,
  author = {},
  title  = {{SentenceTransformer}},
  url    = {https://sbert.net/docs/package_reference/sentence_transformer/SentenceTransformer.html},
  note   = {Accessed: May 15, 2025}
}

@electronic{openwifi,
  author = {},
  title  = {{OpenWiFi}},
  url    = {https://github.com/open-sdr/openwifi},
  note   = {Accessed: Aug 1, 2025}
}

@electronic{RTL8812BU,
  author = {},
  title  = {{RTL8812BU}},
  url    = {https://github.com/fastoe/RTL8812BU},
  note   = {Accessed: Sep 8, 2025}
}

@article{ha2008cubic,
  title={{CUBIC}: a new {TCP-friendly} high-speed {TCP} variant},
  author={Ha, Sangtae and Rhee, Injong and Xu, Lisong},
  journal={ACM SIGOPS operating systems review},
  volume={42},
  number={5},
  pages={64--74},
  year={2008},
  publisher={ACM New York, NY, USA}
}

@article{cui2025towards,
  title={Towards natural language communication for cooperative autonomous driving via self-play},
  author={Cui, Jiaxun and Tang, Chen and Holtz, Jarrett and Nguyen, Janice and Allievi, Alessandro G and Qiu, Hang and Stone, Peter},
  journal={arXiv preprint arXiv:2505.18334},
  year={2025}
}

@inproceedings{gao2025langcoop,
  title={{LangCoop}: Collaborative driving with language},
  author={Gao, Xiangbo and Wu, Yuheng and Wang, Rujia and Liu, Chenxi and Zhou, Yang and Tu, Zhengzhong},
  booktitle={Proceedings of the Computer Vision and Pattern Recognition Conference},
  pages={4226--4237},
  year={2025}
}

@article{bhatt2025uncap,
  title={{UNCAP}: Uncertainty-Guided Planning Using Natural Language Communication for Cooperative Autonomous Vehicles},
  author={Bhatt, Neel P and Li, Po-han and Gupta, Kushagra and Siva, Rohan and Milan, Daniel and Hogue, Alexander T and Chinchali, Sandeep P and Fridovich-Keil, David and Wang, Zhangyang and Topcu, Ufuk},
  journal={arXiv preprint arXiv:2510.12992},
  year={2025}
}

@article{bassamzadeh2024comparative,
  title={A comparative study of {DSL} code generation: Fine-tuning vs. optimized retrieval augmentation},
  author={Bassamzadeh, Nastaran and Methani, Chhaya},
  journal={arXiv preprint arXiv:2407.02742},
  year={2024}
}

@inproceedings{nair2025prompts,
  title={From Prompts to Programs: A {RAG-Based} Framework for Code Synthesis},
  author={Nair, Jaiditya and Kumar, Sunil},
  booktitle={International Conference on Smart Trends for Information Technology and Computer Communications},
  pages={431--438},
  year={2025},
  organization={Springer}
}

@ARTICLE{11264303,
  author={Hu, Yichao and Yang, Furong and Xu, Linlin and Wang, Yuchao and Zhang, Cheng and Song, Chaoyun},
  journal={IEEE Journal of Selected Areas in Sensors}, 
  title={Flexible Near-Field Communication in Wearable Electronics: Antenna Design, Measurements, and System Demonstration}, 
  year={2026},
  volume={3},
  number={},
  pages={23-32},
  doi={10.1109/JSAS.2025.3635540}}

@ARTICLE{7128676,
  author={Chen, Yuh-Shyan and Chiang, Wen-Lin},
  journal={IEEE Sensors Journal}, 
  title={A Spiderweb-Based Massive Access Management Protocol for {M2M} Wireless Networks}, 
  year={2015},
  volume={15},
  number={10},
  pages={5765-5776},
  doi={10.1109/JSEN.2015.2447545}}

@ARTICLE{10693298,
  author={Wang, Xiaonan and Qian, Xinyan},
  journal={IEEE Sensors Journal}, 
  title={Information-Centric {IoT-based} Smart Home Control and Monitoring System}, 
  year={2024},
  volume={24},
  number={21},
  pages={35722-35729},
  doi={10.1109/JSEN.2024.3462929}}

@ARTICLE{10339255,
  author={Chiang, Cheng-Ta},
  journal={IEEE Sensors Journal}, 
  title={A Fish Meat Freshness Detector for {IoT-based} Seafood Market Applications}, 
  year={2024},
  volume={24},
  number={2},
  pages={2049-2054},
  doi={10.1109/JSEN.2023.3335954}}

@ARTICLE{10810266,
  author={Wing Lo, Yuen and Ho Tsoi, Man and Chow, Cheuk-Fai and Mung, Steve W. Y.},
  journal={IEEE Sensors Journal}, 
  title={An {NB-IoT} Monitoring System for Digital Mobile Radio With Industrial {IoT} Performance and Reliability Evaluation}, 
  year={2025},
  volume={25},
  number={3},
  pages={5337-5348},
  doi={10.1109/JSEN.2024.3512859}}

@ARTICLE{4207411,
  author={Depari, Alessandro and Falasconi, Matteo and Flammini, Alessandra and Marioli, Daniele and Rosa, Stefano and Sberveglieri, Giorgio and Taroni, Andrea},
  journal={IEEE Sensors Journal}, 
  title={A New Low-Cost Electronic System to Manage Resistive Sensors for Gas Detection}, 
  year={2007},
  volume={7},
  number={7},
  pages={1073-1077},
  doi={10.1109/JSEN.2007.895965}}

@ARTICLE{10726709,
  author={Zhou, Xiaomao and Hu, Yujiao and Jia, Qingmin and Xie, Renchao},
  journal={IEEE Sensors Journal}, 
  title={Cross-Domain Integration for General Sensor Data Synthesis: Leveraging {LLMs} and Domain-Specific Generative Models in Collaborative Environments}, 
  year={2024},
  volume={24},
  number={24},
  pages={42311-42326},
  doi={10.1109/JSEN.2024.3480932}}

@inproceedings{wang2025cellular,
  title={{Cellular-X}: An LLM-empowered Cellular Agent for Efficient Base Station Operations},
  author={Wang, Liujianfu and Long, Xinyi and Du, Yuyang and Liu, Xiaoyan and Chen, Kexin and Liew, Soung Chang},
  booktitle={Proceedings of the 23rd Annual International Conference on Mobile Systems, Applications and Services},
  pages={625--626},
  year={2025}
}

@article{qi2025verirag,
  title={{VeriRAG}: A Retrieval-Augmented Framework for Automated {RTL} Testability Repair},
  author={Qi, Haomin and Du, Yuyang and Zhang, Lihao and Liew, Soung Chang and Chen, Kexin and Du, Yining},
  journal={arXiv preprint arXiv:2507.15664},
  year={2025}
}

@inproceedings{wang2025rephrase,
  title={Rephrase and Contrast: Fine-Tuning Language Models for Enhanced Understanding of Communication and Computer Networks},
  author={Wang, Liujianfu and Du, Yuyang and Lin, Jingqi and Chen, Kexin and Liew, Soung Chang},
  booktitle={2025 International Conference on Computing, Networking and Communications (ICNC)},
  pages={588--594},
  year={2025},
  organization={IEEE}
}

@article{zhang2025sa,
  title={SA-OOSC: A Multimodal LLM-Distilled Semantic Communication Framework for Enhanced Coding Efficiency with Scenario Understanding},
  author={Zhang, Feifan and Du, Yuyang and Xiang, Yifan and Liu, Xiaoyan and Liew, Soung Chang},
  journal={arXiv preprint arXiv:2509.07436},
  year={2025}
}

@inproceedings{jiao2020openwifi,
  title={{OpenWiFi}: a free and open-source {IEEE802. 11 SDR} implementation on {SoC}},
  author={Jiao, Xianjun and Liu, Wei and Mehari, Michael and Aslam, Muhammad and Moerman, Ingrid},
  booktitle={2020 IEEE 91st vehicular technology conference (VTC2020-Spring)},
  pages={1--2},
  year={2020},
  organization={IEEE}
}

@article{sun2022recent,
  title={Recent advances in {LoRa}: A comprehensive survey},
  author={Sun, Zehua and Yang, Huanqi and Liu, Kai and Yin, Zhimeng and Li, Zhenjiang and Xu, Weitao},
  journal={ACM Transactions on Sensor Networks},
  volume={18},
  number={4},
  pages={1--44},
  year={2022},
  publisher={ACM New York, NY}
}

@article{yang2025qwen3,
  title={Qwen3 technical report},
  author={Yang, An and Li, Anfeng and Yang, Baosong and Zhang, Beichen and Hui, Binyuan and Zheng, Bo and Yu, Bowen and Gao, Chang and Huang, Chengen and Lv, Chenxu and others},
  journal={arXiv preprint arXiv:2505.09388},
  year={2025}
}

@inproceedings{hsieh2023distilling,
  title={Distilling step-by-step! outperforming larger language models with less training data and smaller model sizes},
  author={Hsieh, Cheng-Yu and Li, Chun-Liang and Yeh, Chih-Kuan and Nakhost, Hootan and Fujii, Yasuhisa and Ratner, Alex and Krishna, Ranjay and Lee, Chen-Yu and Pfister, Tomas},
  booktitle={Findings of the Association for Computational Linguistics: ACL 2023},
  pages={8003--8017},
  year={2023}
}

@article{ouyang2022training,
  title={Training language models to follow instructions with human feedback},
  author={Ouyang, Long and Wu, Jeffrey and Jiang, Xu and Almeida, Diogo and Wainwright, Carroll and Mishkin, Pamela and Zhang, Chong and Agarwal, Sandhini and Slama, Katarina and Ray, Alex and others},
  journal={Advances in neural information processing systems},
  volume={35},
  pages={27730--27744},
  year={2022}
}

\end{document}